\documentclass[twocolumn]{aastex631}

\usepackage{longtable}
\usepackage{amsmath}
\usepackage{makecell}
\usepackage{color}
\usepackage{soul}
\soulregister{\cite}7
\soulregister{\citep}7

\begin{document}
\title{Strong [O {\footnotesize \uppercase\expandafter{\romannumeral3}}] $\lambda $5007 Compact Galaxies Identified from SDSS DR16 and Their Scaling Relations}

\author{Weiyu Ding}
\affiliation{Deep Space Exploration Laboratory / Department of Astronomy, University of Science and Technology of China, Hefei 230026, China}
\affiliation{School of Astronomy and Space Science, University of Science and Technology of China, Hefei 230026, China}
\affiliation{Key Laboratory of Optical Astronomy, National Astronomical Observatories, Chinese Academy of Sciences, Beijing 100012, China}

\author{Hu Zou}
\affiliation{Key Laboratory of Optical Astronomy, National Astronomical Observatories, Chinese Academy of Sciences, Beijing 100012, China}

\author{Xu Kong}
\affiliation{Deep Space Exploration Laboratory / Department of Astronomy, University of Science and Technology of China, Hefei 230026, China}
\affiliation{School of Astronomy and Space Science, University of Science and Technology of China, Hefei 230026, China}

\author{Yulong Gao}
\affiliation{School of Astronomy and Space Science, Nanjing University, Nanjing 210093, China}              
\affiliation{Key Laboratory of Modern Astronomy and Astrophysics (Nanjing University), Ministry of Education, Nanjing 210093, China}

\author{Fujia Li}
\affiliation{Deep Space Exploration Laboratory / Department of Astronomy, University of Science and Technology of China, Hefei 230026, China}
\affiliation{School of Astronomy and Space Science, University of Science and Technology of China, Hefei 230026, China}

\author{Hongxin Zhang}
\affiliation{Department of Astronomy, University of Science and Technology of China, Hefei 230026, China}
\affiliation{School of Astronomy and Space Science, University of Science and Technology of China, Hefei 230026, China}

\author{Jiali Wang}
\affiliation{Key Laboratory of Optical Astronomy, National Astronomical Observatories, Chinese Academy of Sciences, Beijing 100012, China}

\author{Jie Song}
\affiliation{Deep Space Exploration Laboratory / Department of Astronomy, University of Science and Technology of China, Hefei 230026, China}
\affiliation{School of Astronomy and Space Science, University of Science and Technology of China, Hefei 230026, China}

\author{Jipeng Sui}
\affiliation{Key Laboratory of Optical Astronomy, National Astronomical Observatories, Chinese Academy of Sciences, Beijing 100012, China}

\author{Jundan Nie}
\affiliation{Key Laboratory of Optical Astronomy, National Astronomical Observatories, Chinese Academy of Sciences, Beijing 100012, China}

\author{Suijian Xue}
\affiliation{Key Laboratory of Optical Astronomy, National Astronomical Observatories, Chinese Academy of Sciences, Beijing 100012, China}

\author{Weijian Guo}
\affiliation{Key Laboratory of Optical Astronomy, National Astronomical Observatories, Chinese Academy of Sciences, Beijing 100012, China}

\author{Yao Yao}
\affiliation{Deep Space Exploration Laboratory / Department of Astronomy, University of Science and Technology of China, Hefei 230026, China}
\affiliation{School of Astronomy and Space Science, University of Science and Technology of China, Hefei 230026, China}

\author{Zhimin Zhou}
\affiliation{Key Laboratory of Optical Astronomy, National Astronomical Observatories, Chinese Academy of Sciences, Beijing 100012, China}

\correspondingauthor{Hu Zou, Xu Kong}
\email{zouhu@nao.cas.cn, xkong@ustc.edu.cn}



\begin{abstract}
Green pea galaxies are a special class of star-forming compact galaxies with strong [O {\footnotesize \uppercase\expandafter{\romannumeral3}}]$\lambda$5007 and considered as analogs of high-redshift Ly$\alpha$-emitting galaxies and potential sources for cosmic reionization. In this paper, we identify 76 strong [O {\footnotesize \uppercase\expandafter{\romannumeral3}}]$\lambda$5007 compact galaxies at $z < 0.35$ from DR16 of the Sloan Digital Sky Survey. These galaxies present relatively low stellar mass, high star formation rate, and low metallicity. Both star-forming main sequence relation (SFMS) and mass-metallicity relation (MZR) are investigated and compared with green pea and blueberry galaxies collected from literature. It is found that our strong [O {\footnotesize \uppercase\expandafter{\romannumeral3}}] $\lambda$5007 compact galaxies share common properties with those compact galaxies with extreme star formation and show distinct scaling relations in respect to those of normal star-forming galaxies at the same redshift. The slope of SFMS is higher, indicates that strong [O {\footnotesize \uppercase\expandafter{\romannumeral3}}]$\lambda$5007 compact galaxies might grow faster in stellar mass. The lower MZR implies that they may be less chemically evolved and hence on the early stage of star formation. A further environmental investigation confirms that they inhabit relatively low-density regions. Future large-scale spectroscopic surveys will provide more details on their physical origin and evolution.
\end{abstract}



\section{INTRODUCTION} \label{sec:intro}
In modern extragalactic astrophysics, one of the most significant topics is to understand main physical processes occurring during the epoch of reionization (EoR). In the early universe, the intergalactic medium (IGM) changed from being neutral and opaque to ionized and transparent. Reionization seems to have been completed by $z\sim 6$ \citep{2006AJ....132..117F,2015MNRAS.447..499M}. Star-forming galaxies \citep[SFGs; e.g.,][]{2010Natur.468...49R, 2019ApJ...879...36F, 2020MNRAS.496.4574Y}, active galactic nuclei \citep[AGN; e.g.,][]{1998ApJ...503..505H,2021MNRAS.508.2706Y,2022ApJ...938...25F}, and quasars \citep[e.g.,][]{2004ApJ...604..484M} are possible candidates for the source of the ionizing radiation. It is generally believed that star-forming galaxies are the main contributors to reionization. However, star-forming regions typically have a large volume of HI gas around them, which prevents the ionizing radiation emitted by hot stars from escaping into the IGM. If low and intermediate mass galaxies have a fully-ionized interstellar medium or galaxies could be perforated by optically thin tunnels so that the ionizing radiation could escape, they may get around this problem \cite[]{2014MNRAS.442..900N,2015ApJ...805...14R}. Although significant progress has been made over the past 5 to 10 years, the production and escape of ionizing radiation in star-forming galaxies have not been fully understood yet.

Due to the difficulty of observing high-redshift galaxies that leak Lyman-continuum radiation (LyC, $\lambda_\mathrm{}{rest} < 912\mathrm{\AA}$), it is critical to find some low-redshift analogies. Green pea (GP) galaxies have many similarities with high-redshift SFGs, such as high specific star formation rates and low metallicities \citep{2011ApJ...728..161I,2014MNRAS.442..900N,2015ApJ...809...19H}. A significant proportion of GP galaxies show high LyC escape fractions ranging from 1-50\% \citep{2015ApJ...809...19H,2017A&A...597A..13V,2020MNRAS.491..468I}. It makes GP galaxies crucial systems for studying the escape mechanism of ionizing radiation.

Green pea galaxies were originally identified in the Galaxy Zoo project \citep{2008MNRAS.389.1179L} by citizen scientists. Following this discovery, \cite{2009MNRAS.399.1191C} used unique characteristics of compact size and green color caused by the strong [O {\footnotesize \uppercase\expandafter{\romannumeral3}}]$\lambda$5007 emission line to discover the first sample of GP galaxies from the Sloan Digital Sky Survey \citep[SDSS;][]{2000AJ....120.1579Y}. A total of 251 GP galaxies from SDSS DR7 are discovered by color selection in the redshift range of $0.112 < z < 0.360$, among which only 80 of them have optical spectra. Different from the typical GP definition, \citet{2019ApJ...872..145J} used the equivalent width of either [O {\footnotesize \uppercase\expandafter{\romannumeral3}}]$\lambda$5007 or H$\beta$ to obtain 800 GP-like galaxies from SDSS DR13. \citet{2020ApJ...898...68B} used strong [O {\footnotesize \uppercase\expandafter{\romannumeral3}}]$\lambda$5007 emission lines to find 13 GP galaxies in the redshift range of $0.29 < z < 0.41$ from the KPNO International Spectroscopic Survey \citep[KISS;][]{2000AAS...197.7606S}. Generally, green pea galaxies are selected in a relatively narrow redshift range. However, blueberry and purple grape galaxies are thought to be essentially identical to GP galaxies, except that they are located in different redshift ranges \citep{2022ApJ...927...57L}. Blueberry galaxies are young starburst galaxies at $z < 0.05$ \citep{2017ApJ...847...38Y}. In this redshift range, the [O {\footnotesize \uppercase\expandafter{\romannumeral3}}]$\lambda$5007 emission line is within $g$ band, which makes their colors blue. As a contrast, GP galaxies in the redshift range of $0.112 < z < 0.36$ have the [O {\footnotesize \uppercase\expandafter{\romannumeral3}}]$\lambda$5007 line within $r$ band. Purple grape galaxies are those at $z > 0.36$ with [O {\footnotesize \uppercase\expandafter{\romannumeral3}}]$\lambda$5007 within $i$ band and the UV continuum redshifted to $g$ band and are those at $0.05< z < 0.112$ with both strong [O {\footnotesize \uppercase\expandafter{\romannumeral3}}]$\lambda$5007 within $g$ band and H$\alpha$ within $i$ band.

\begin{figure*}[htbp!]
\includegraphics[width=\linewidth]{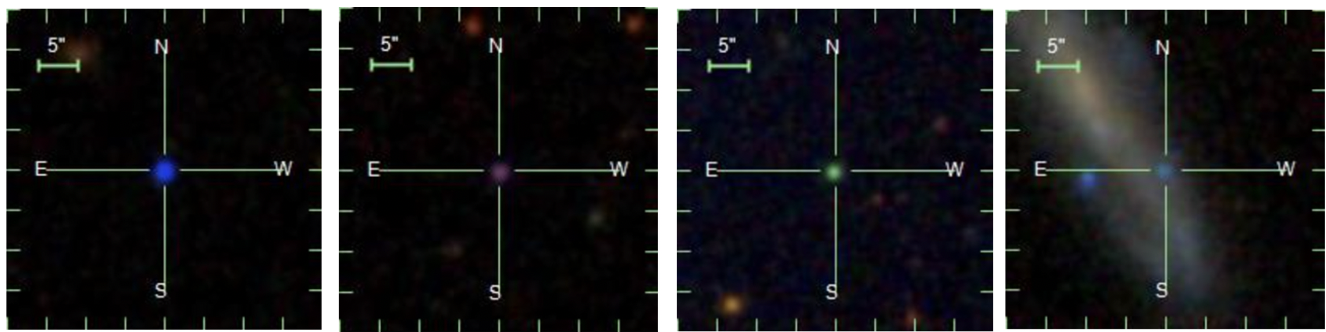}
\caption{SDSS color images of some galaxies in our selected sample. The first three galaxies are GP-like compact galaxies at $z = $0.04, 0.09, and 0.22, respectively. The last panel shows a galaxy excluded by us through visual examination, which might be a star formation region of a large galaxy. 
\label{fig:colorimage}}
\end{figure*}

In this paper, we intend to compile a catalog of galaxies with strong [O {\footnotesize \uppercase\expandafter{\romannumeral3}}]$\lambda$5007 emission lines from SDSS DR16 \citep{2020ApJS..249....3A} and explore their physical properties and scaling relations. The paper is organized as follows. In Section \ref{sec:data}, we describe the data and sample selection criteria. In Section \ref{sec:properties}, we describe the measurements of physical properties for the galaxy sample. Section \ref{sec:result} presents the basic properties and scaling relations including SFMS and MZR. Comparisons with other GP galaxies and GP-like galaxies from the literature are also shown in this section. Section \ref{sec:summary} gives a summary. In this paper, we assume a $\Lambda$-CDM standard cosmology with $H_{0} = 70$ km s$^{-1}$ Mpc$^{-1}$, $\Omega_{m} = 0.3$, and $\Omega_{\Lambda} = 0.7$. All photometric magnitudes are in AB mag.

\section{Data and Sample} \label{sec:data}
\subsection{Spectroscopic data and data preprocessing}
The spectroscopic data used in this paper are from the sixteenth data release \citep[DR16;][]{2020ApJS..249....3A} of the Sloan Digital Sky Survey \citep[SDSS;][]{2000AJ....120.1579Y}. DR16 includes the data of the extended Baryon Oscillation Spectroscopic Survey \citep[eBOSS;][]{2016AJ....151...44D} of SDSS IV \citep{2017AJ....154...28B}. The survey uses the Sloan Foundation 2.5m Telescope at Apache Point Observatory \citep{2006AJ....131.2332G} and the BOSS spectrograph \citep{2013AJ....145...10D}, which has 1000 fibers per 7-deg$^2$ plate at a resolution of $\rm{R}\sim2000$ and covers the wavelength range of 3600--10000 {\AA}. The eBOSS aims to study the expansion history of the universe by using different tracers of spectroscopic reshifts, including luminous red galaxies, emission line galaxies, and quasars. There are a total of 1.4 million spectra observed by the eBOSS \citep{2020ApJS..249....3A}, where about 860,000 galaxies and quasars are new with respect to previous data releases.

We choose the objects in DR16 with the spectroscopic classification of "GALAXY" and reliable redshift measurements. The spectral fitting code of STARLIGHT \citep{{2005MNRAS.358..363C}} is used to derive the underlying stellar continuum for each galaxy. It is subtracted from the observed spectrum in order to measure intrinsic fluxes of emission lines generated from the gas. In STARLIGHT, we  use single stellar populations (SSPs) from \cite{2003MNRAS.344.1000B} models, the \textbf{\textcolor{blue}{\cite{1955ApJ...121..161S}}} initial mass function (IMF) and the attenuation law of \cite{2000ApJ...533..682C}. The fluxes of strong emission lines (e.g., H$\beta$, [O {\footnotesize \uppercase\expandafter{\romannumeral3}}]$\lambda$$\lambda$4959,5007, H$\alpha$, and [N II]$\lambda$6583) are calculated by the Gaussian-profile fitting with an IDL package MPFIT \citep{2009ASPC..411..251M}. Following the method of \cite{2014ApJ...780..122L}, we also estimate the flux errors and signal-to-noise ratios (S/N). The gas-phase extinction is calculated using the Balmer decrement with the assumption of the intrinsic flux ratio $(\rm{H}\alpha/\rm{H}\beta)_0$ = 2.86 under the case-B recombination. All the fluxes of emission lines are extinction-corrected with the \cite{2000ApJ...533..682C} attenuation curve.

\subsection{Sample selection} \label{sample selection}
The strong [O {\footnotesize \uppercase\expandafter{\romannumeral3}}]$\lambda$5007 compact galaxies are selected using the following criteria: 
\begin{itemize}
\item[1] [O {\footnotesize \uppercase\expandafter{\romannumeral3}}]$\lambda$5007 line is detected with flux S/N greater than 5;
\item[2] [O {\footnotesize \uppercase\expandafter{\romannumeral3}}]$\lambda$5007 line is strong enough, where its equivalent width is larger than 200 {\AA} \citep{2021ApJ...909...52M};
\item[3] The galaxy has a compact size, i.e., the radius containing 90\% of the Petrosian flux in SDSS $r$ band (PetroR90\_r) is smaller than 3\arcsec \citep{2019ApJ...872..145J}.
\end{itemize}

We obtain a total of 162 galaxies satisfying the above selection criteria in a redshift range of  $z < 0.48$. These galaxies are visually examined with SDSS composite color images. It is found that quite a few galaxies appear to be a part of large galaxies (possibly HII regions, e.g. the rightmost panel of Figure \ref{fig:colorimage}) or contaminated by neighbours. Their photometry and corresponding properties could be considerably affected by host or nearby galaxies. A total of 85 galaxies are discarded.

\begin{figure}[htbp!]
\centering
\includegraphics [width=\linewidth]{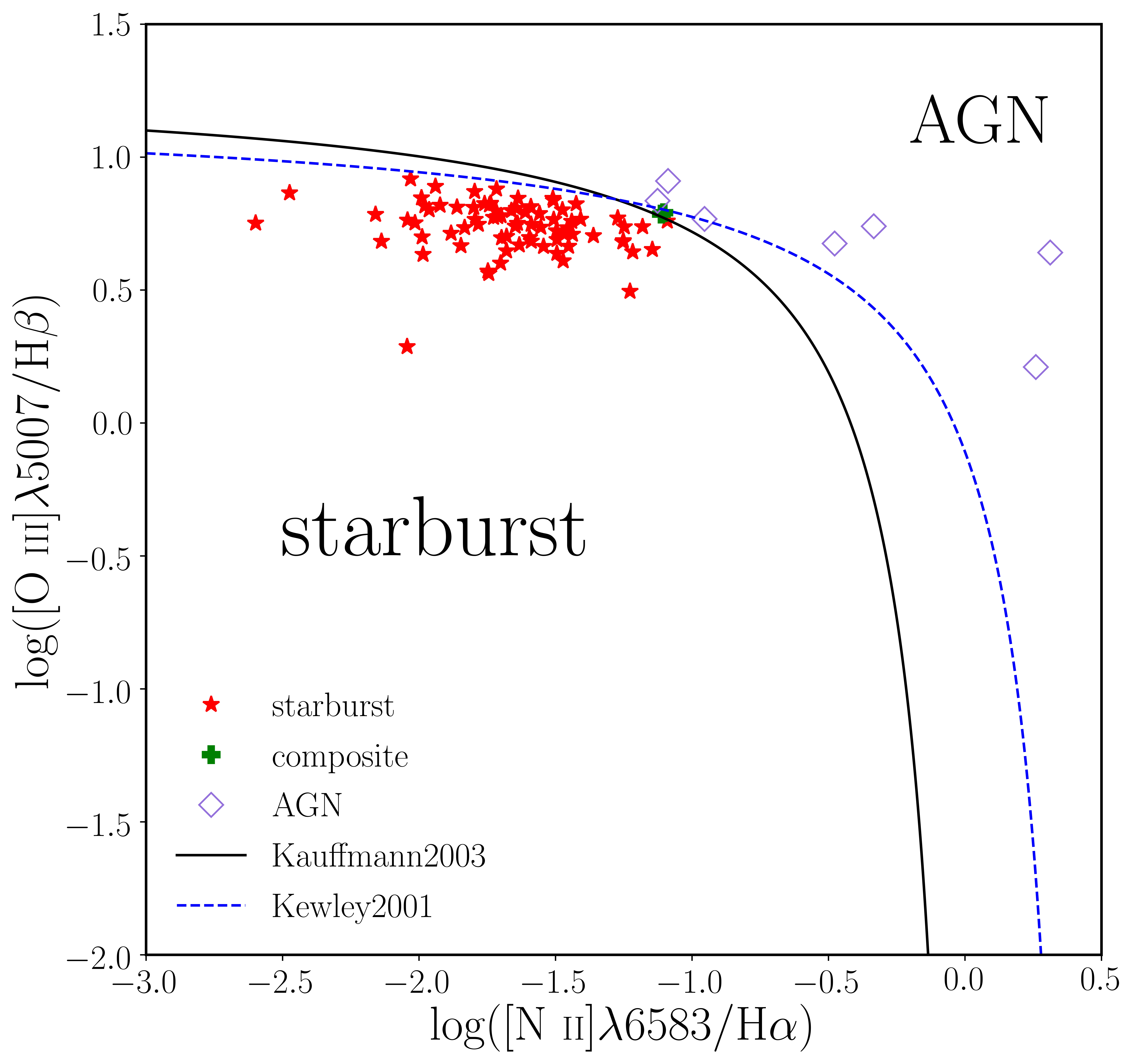}
\caption{Optical diagnostic diagram (BPT diagram) separating our strong [O {\footnotesize \uppercase\expandafter{\romannumeral3}}]$\lambda$5007 compact galaxies into star-forming galaxies in red stars, composite galaxies in green crosses, and AGNs in purple diamonds. The dashed curve is the theoretical separation limit for star-forming galaxies proposed by \cite{2001ApJ...556..121K}, while the solid curve is the classification lower limit for selecting AGNs proposed by \cite{2003MNRAS.346.1055K}. 
\label{fig:bpt}}
\end{figure}

To exclude galaxies with AGN activities, we apply the BPT diagram \citep{Baldwin_1981,1987ApJS...63..295V} to classify our galaxies into different spectral types according to the discrimination lines proposed by \cite{2001ApJ...556..121K} and \cite{2003MNRAS.346.1055K}. The BPT diagram is shown in Figure \ref{fig:bpt}. There are 76 galaxies classified as star-forming galaxies (red pentagrams), 2 galaxies as composite (green crosses), and 7 galaxies as AGNs (purple diamonds). We are only concerned with star-forming galaxies, so AGNs are not analyzed in the rest of this paper. The clustering of galaxies in the top left of Figure \ref{fig:bpt} indicates that our star-forming galaxies have high ionization intensity ratios and possibly have low metallicities. 

\begin{figure*}[htbp!]
\includegraphics [width=\linewidth]{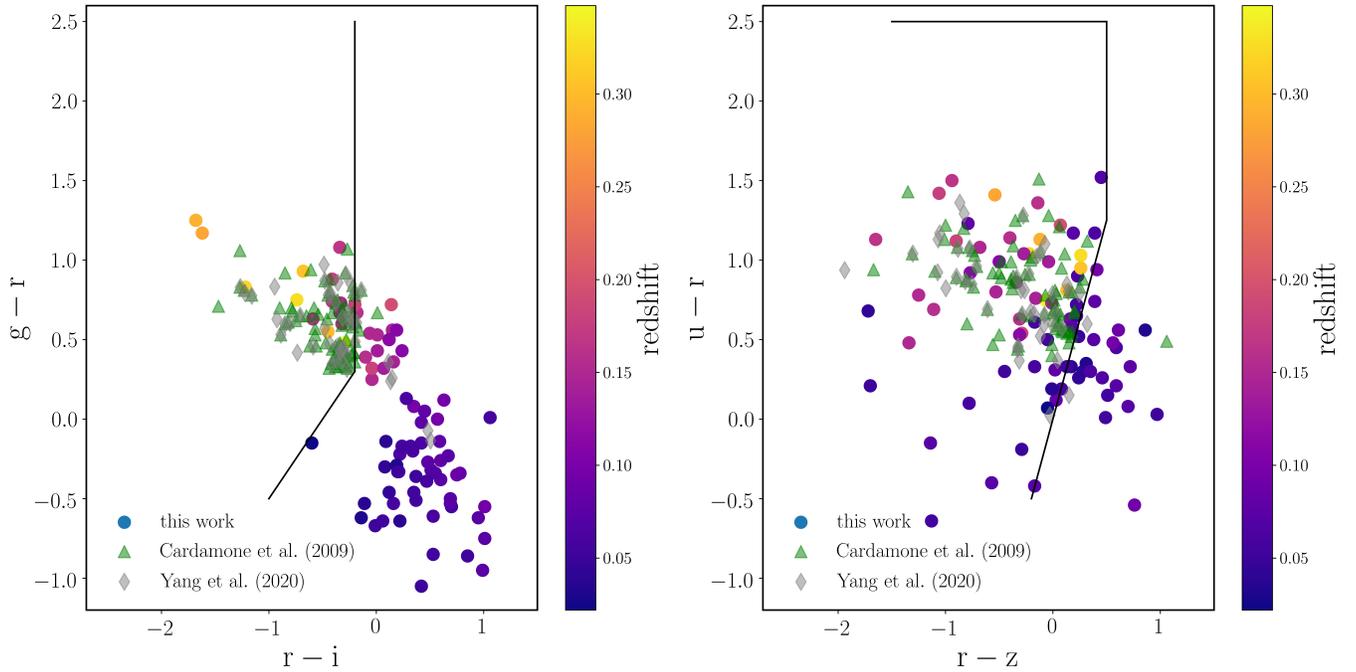}
\caption{Left: color-color diagram of $r - i$ versus $g - r$ for our galaxy sample colored by redshift. Right: color-color diagram of $r - z$ versus $u - r$. Green pea galaxies from \cite{2009MNRAS.399.1191C} (green triangles) and \cite{2017ApJ...847...38Y} (grey diamonds) are also overplotted. The solid lines mark the color selection criteria in \cite{2009MNRAS.399.1191C}. 
\label{fig:colorspace}}
\end{figure*}

We have finally chosen 76 galaxies in the redshift range of $0.02 < z < 0.35$. The first three panels in Figure \ref{fig:colorimage} display the color images of three typical examples at different redshifts in our final sample. Figure \ref{fig:colorspace} shows the galaxy distributions on two color-color diagrams. In this figure, we also show the green pea galaxies from \cite{2009MNRAS.399.1191C} at $0.112 < z < 0.360$ in yellow triangles, and those from \cite{2017ApJ...847...38Y} at $0.098 < z < 0.34$ in grey diamonds. The black solid lines mark the color selection criteria from \cite{2009MNRAS.399.1191C}. Many of our strong [O {\footnotesize \uppercase\expandafter{\romannumeral3}}]$\lambda$5007 compact galaxies are located out of the color boxes proposed by \cite{2009MNRAS.399.1191C}. Our galaxies have lower redshifts but approximate [O {\footnotesize \uppercase\expandafter{\romannumeral3}}] equivalent widths and sizes of GP galaxies, so they should be the local counterparts of such galaxies. The galaxy distribution in the color space could be used for defining new selection criteria of GP-like galaxies at $z< 0.35$.  

\section{Physical properties of our strong [O {\footnotesize \uppercase\expandafter{\romannumeral3}}]$\lambda$5007 compact galaxies} \label{sec:properties}

\subsection{Star formation rate} \label{subsec:SFR}
Strong emission lines are usually used to estimate the star formation rate (SFR). Due to low redshifts of our galaxies, we adopt the most accurate SFR calibration based on the H$\alpha$ luminosity to estimate SFR \citep{1998ARA&A..36..189K}:
\begin{equation}
    \mathrm{SFR}\left(\mathrm{M}_{\odot} \mathrm{yr}^{-1}\right)=7.9 \times 10^{-42} \, L_\mathrm{H \alpha}\left(\mathrm{erg} \mathrm{s}^{-1}\right)
\end{equation}
where the H$\alpha$ luminosity ($L_\mathrm{H\alpha}$) is calculated with the extinction-corrected H$\alpha$ flux and the luminosity distance from the flat $\Lambda$-CDM cosmology. The gas-phase extinction is derived through the Balmer decrement with the assumption of the intrinsic Balmer line ratio of $(\mathrm{H\alpha}/\mathrm{H\beta})_0 = 2.86$ and the \cite{2000ApJ...533..682C} reddening curve. The median gas-phase extinction E(B-V) of our galaxies is about 0.11 mag. The median SFR is about 2.81 $\mathrm{M_{\odot} yr^{-1}}$ with highest value up to 18.38 $\rm{M_{\odot} yr^{-1}}$, suggesting that our strong [O {\footnotesize \uppercase\expandafter{\romannumeral3}}]$\lambda$5007 compact galaxies have strong star formation.


\subsection{Stellar mass} \label{subsec:mass}
\begin{figure}[ht!]
\includegraphics [width=\linewidth]{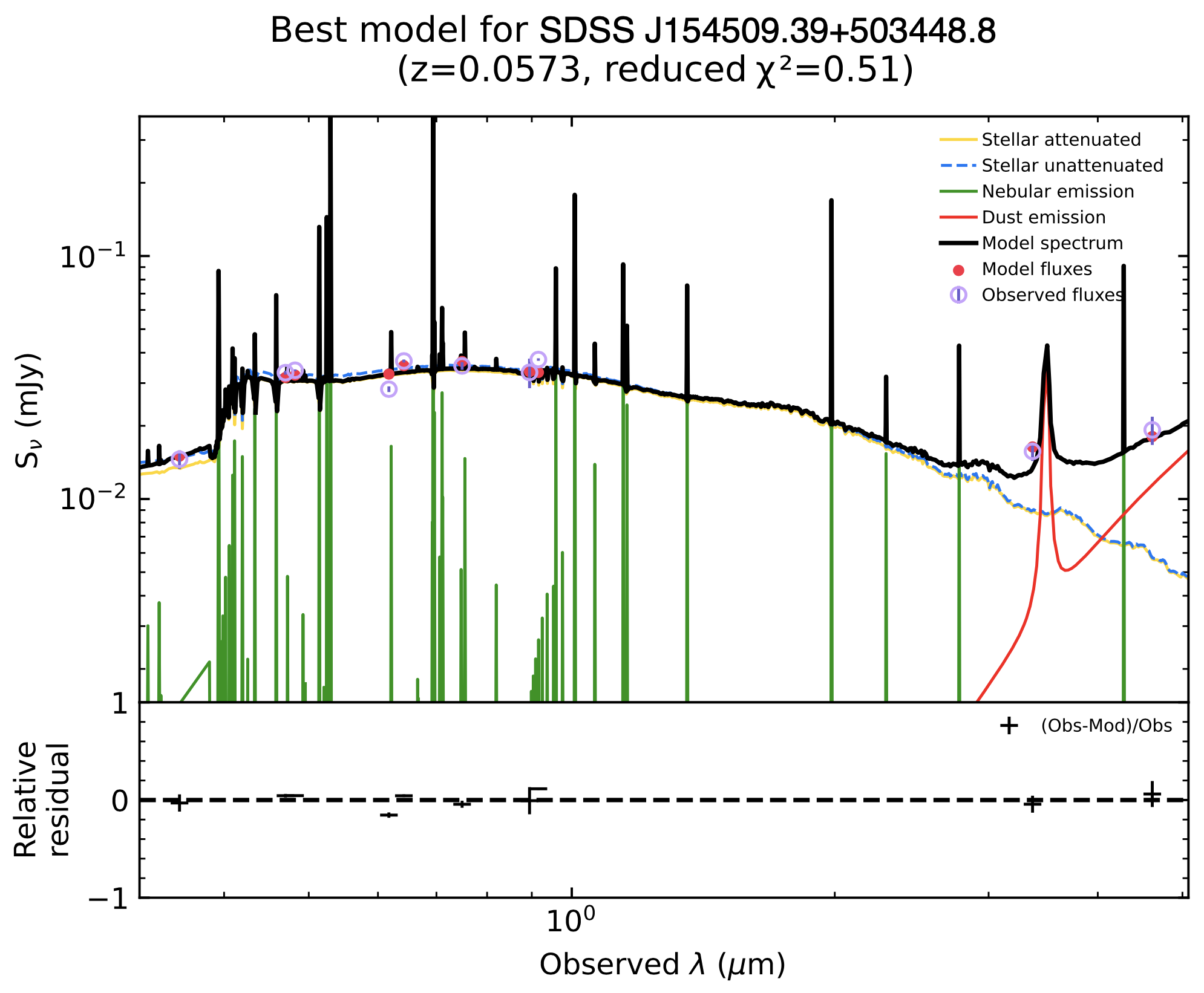}
\caption{A SED-fitting example using CIGALE with the SDSS $ugriz$ DESI $grz$ and WISE $W1,W2$ photometry. This plot is generated by CIGALE, where the upper panel shows the models and photometric data and the bottom panel presents the residual between the observed and model data.\label{fig:cigale}}
\end{figure}

The stellar mass is estimated by fitting multi-wavelength spectral energy distributions (SEDs) with the stellar population synthesis code of CIGALE \citep{2005MNRAS.360.1413B,2009A&A...507.1793N,2019A&A...622A.103B}. The photometric data include the photometric magnitudes of SDSS $ugriz$, DESI $grz$, and WISE $W1W2$. In CIGALE, we adopt the delayed-$\tau$ star formation history, BC03 \citep{2003MNRAS.344.1000B} stellar population models, Salpeter IMF\citep{1955ApJ...121..161S}, nebular emission-line models, \cite{2000ApJ...533..682C} reddening curve and dust emission model from \citet{2014ApJ...780..172D}. An example of the CIGALE SED fitting (SDSS J154509.39+503448.8)is shown in Figure \ref{fig:cigale}. The stellar mass of our galaxy sample ranges from $7 \times 10^{5}$ $\rm{M_\odot}$ to $3 \times 10^{9}$ $\rm{M_\odot}$ and the median value is $5 \times 10^{7}$ $\rm{M_\odot}$. Most galaxies are low-mass dwarf galaxies.


\subsection{Metallicity} \label{subsec:Metallicity}
There are two widely-used methods to measure the gas-phase metallicity of galaxies: one is based on the electron temperature ($T_e$) \citep{1984ASSL..112.....A,2018ApJ...869...15G,2022ApJ...926...57Y} and the other is based on emission-line ratios of strong emission lines \citep{2004ApJ...617..240K,2004MNRAS.348L..59P,2008A&A...488..463M}. The direct $T_e$ method relies on the detection of [O {\footnotesize \uppercase\expandafter{\romannumeral3}}]$\lambda$4363, which is undetectable in most of our galaxy spectra. We choose N2-based method described in \cite{2004MNRAS.348L..59P}, which is suitable for galaxies with low metallicity:
\begin{eqnarray}
12+\log (\mathrm{O} / \mathrm{H}) &= 8.90+0.57 \times \mathrm{N}2,  \nonumber \\
\mathrm{N}2 &\equiv \log\left(\frac{[\mathrm{N {\scriptstyle II}}]{\lambda}6583}{\mathrm{H}\alpha}\right)\\
\end{eqnarray}
where the calibration is applicable for $-2.5<\mathrm{N}2<-0.3$. All our galaxies have N2 within this range. The overall metallicity for our galaxy sample is about $\rm{log[O/H] + 12 \sim 7.96}$.  Table \ref{tab:Properties} lists all the properties of our strong [O {\footnotesize \uppercase\expandafter{\romannumeral3}}]$\lambda$5007 compact galaxies. Note that the errors of SFR and metallicity in this table are propagated from the uncertainties of emission-line measurements, while the error of stellar mass is provided by the CIGALE SED fitting.

\section{Results and discussions} \label{sec:result}
\subsection{EW([O {\footnotesize \uppercase\expandafter{\romannumeral3}}]) vs. SFR} \label{subsec:EW-SFR}
\begin{figure}[htbp]
\centering
\includegraphics [width=\linewidth]{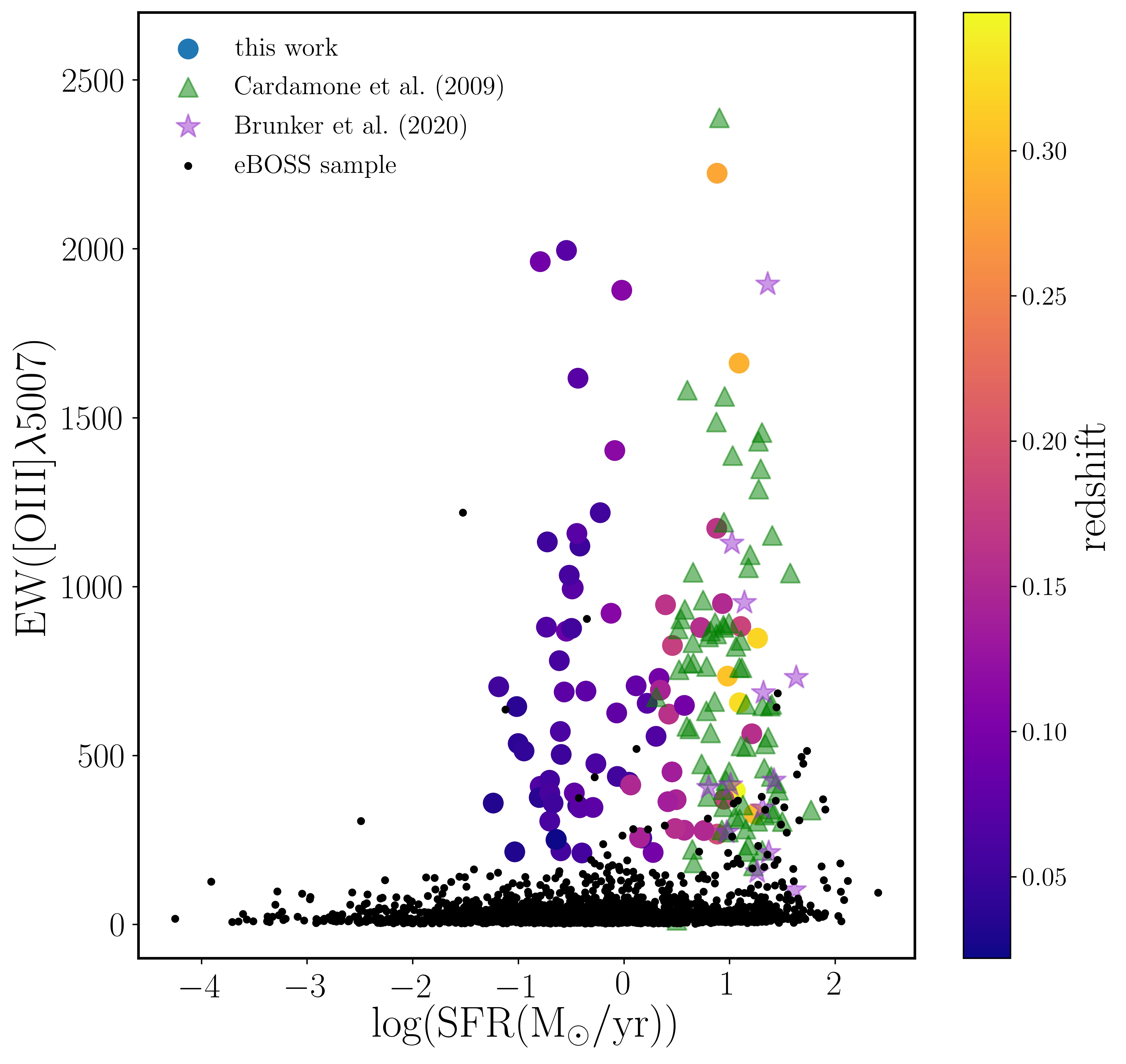}
\caption{[O {\footnotesize \uppercase\expandafter{\romannumeral3}}]$\lambda$5007 equivalent width against SFR colorcoded by redshift. Green pea galaxies from \cite{2009MNRAS.399.1191C} (\textbf{\textcolor{blue}{green triangles}}), \cite{2020ApJ...898...68B} (purple stars),  and eBOSS H$\alpha$-detected star-forming galaxies (black dots) are also plotted. \label{fig:EWSFR}}
\end{figure}

Figure \ref{fig:EWSFR} shows the equivalent widths of [O {\footnotesize \uppercase\expandafter{\romannumeral3}}]$\lambda$5007 versus SFR of our galaxies. The plot is color-coded by redshift. In this plot, we also present the green-pea galaxies of \citet{2009MNRAS.399.1191C} and \citet{2020ApJ...898...68B} for comparisons. To exhibit the difference between the strong [O {\footnotesize \uppercase\expandafter{\romannumeral3}}]$\lambda$5007 compact galaxies and normal star-forming galaxies, we randomly select 2000 H$\alpha$-detected galaxies with S/N $>$ 5 from eBOSS whose redshifts are in the same range of our sample. From Figure \ref{fig:EWSFR}, we can see that our strong [O {\footnotesize \uppercase\expandafter{\romannumeral3}}]$\lambda$5007 galaxies at $0.112 < z < 0.36$ occupy the same region as green-pea galaxies of \cite{2009MNRAS.399.1191C} and \cite{2020ApJ...898...68B}, whose SFR is around 14 $M_\odot$ yr$^{-1}$. Our galaxies at lower redshift present lower SFRs, which is about 24 times less than green pea galaxies. They are low-redswhift counterparts of green pea galaxies. Meanwhile, there is almost no overlap between our and H$\alpha$-detected galaxies, indicates that the strong [O {\footnotesize \uppercase\expandafter{\romannumeral3}}]$\lambda$5007 compact galaxies are extremely rare in the low-redshift universe.

\subsection{Star forming main sequence} \label{subsec:M-SFR}
\begin{figure}[htbp]
\centering
\includegraphics [width=\linewidth]{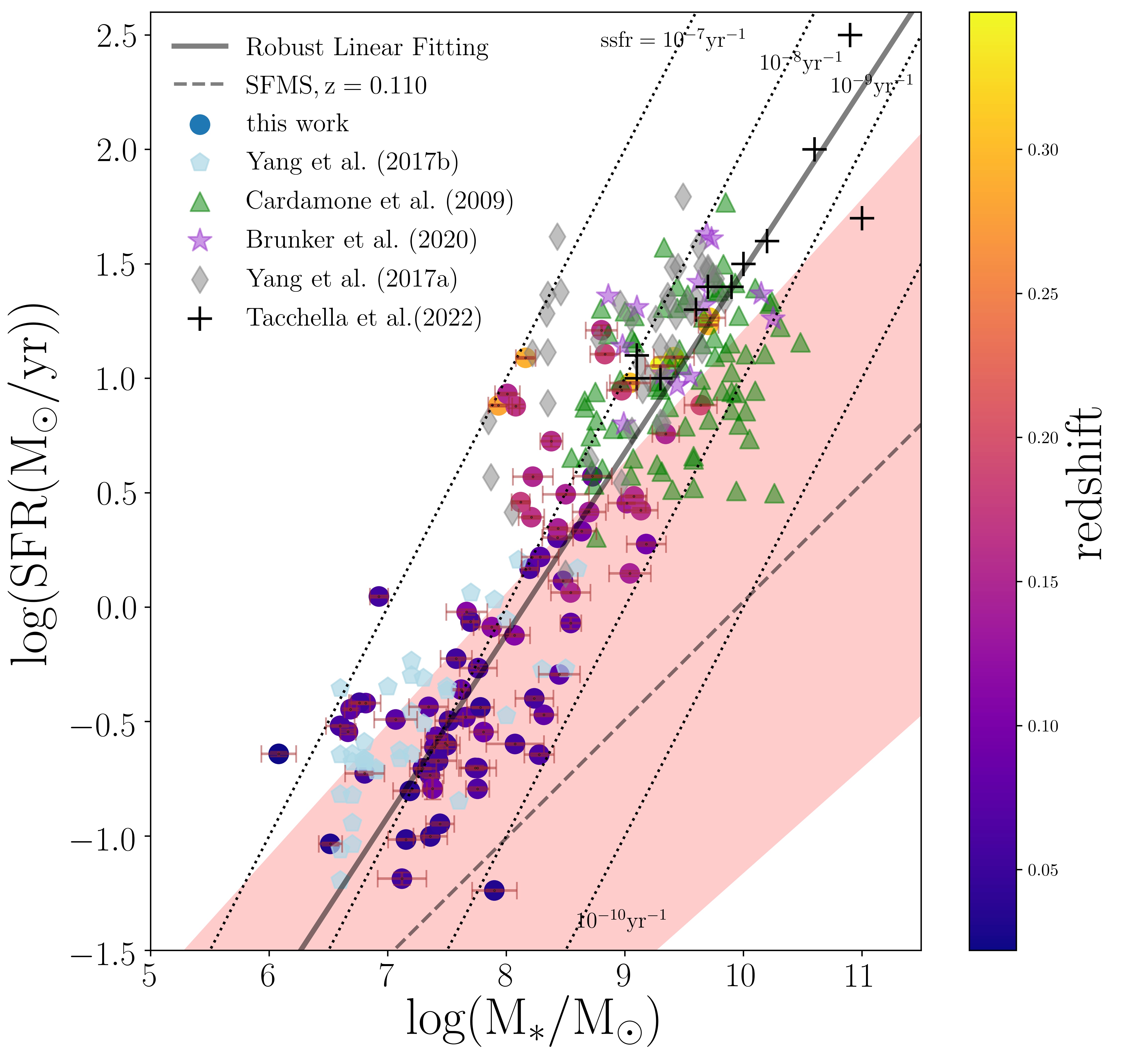}
\caption{Relation between stellar mass and SFR for our strong [O {\footnotesize \uppercase\expandafter{\romannumeral3}}]$\lambda$5007 compact galaxies (filled circles) color-coded by redshift. The error bars in orange show the measurement uncertainties. Blueberry galaxies (light blue pentagons) from \cite{2017ApJ...844..171Y}, green pea galaxies from \cite{2009MNRAS.399.1191C} (green triangles), \cite{2020ApJ...898...68B} (purple stars), \cite{2017ApJ...847...38Y} (grey diamonds) and \textbf{\textcolor{blue}{\cite{2022ApJ...927..170T} (black cross)}}are also overplotted. The star-forming main sequence for normal galaxies at redshift of $z=0.11$ from \cite{2014ApJS..214...15S} is plotted in dashed line for comparison and the red shade indicates the uncertainty of this SFMS. The solid line represents the best robust linear fitting to our galaxy sample. The dotted lines present the constant sSFRs of 10$^{-7}$, 10$^{-8}$, 10$^{-9}$, and 10$^{-10}$ yr$^{-1}$.
\label{fig:MSFR}}
\end{figure}

Figure \ref{fig:MSFR} demonstrates the relation between the stellar mass and SFR, which is also known as star forming main sequence \citep[SFMS;][]{2014ApJS..214...15S}. The SFR increases steadily with the redshift.  In this figure, we also present blueberry galaxies from \cite{2017ApJ...844..171Y}, green pea galaxies from \cite{2009MNRAS.399.1191C}, \cite{2017ApJ...847...38Y} and \cite{2020ApJ...898...68B}. It can be seen that our galaxy sample covers the whole main sequence of both green pea and blueberry galaxies. They seem to lie in the same sequence, confirming that they are likely to belong to the same type of galaxies but at different redshifts. The robust linear fitting to the main sequence of our galaxies is expressed as: 

\begin{equation}
\log(\mathrm{SFR})=0.79 \log{M_{*}}-6.48.
\end{equation}
We also perform a robust linear fitting to all galaxies shown in Figure \ref{fig:MSFR}, giving {$\log(\mathrm{SFR})=0.78 \log {M_{*}}-6.21$}. There is no significant difference between these two linear fittings.  

We also compare our main sequence relation with the one for normal star forming galaxies from \cite{2014ApJS..214...15S}, which is expressed as:
\begin{equation}
    \begin{aligned}
        \log \mathrm{SFR}\left({M}_{*}, t \right) &= (0.84 \pm 0.02-0.026 \pm 0.003 t) \log{{M_{*}}} \\ 
        &-(6.51 \pm 0.24-0.11 \pm 0.03 t),
    \end{aligned}
\end{equation}
where $t$ is the age of the universe in Gyr. The normal main sequence at $z = 0.11$ (the median redshift of our sample) is overplotted in Figure \ref{fig:MSFR}. It is evident that our galaxies show much higher SFR at specified stellar mass. Similar to the results in \cite{2009MNRAS.399.1191C} and \cite{2022ApJ...927...57L}, our strong strong [O {\footnotesize \uppercase\expandafter{\romannumeral3}}]$\lambda$5007 galaxies also have much higher specific SFR (sSFR). In addition, the main-sequence slope of normal galaxies \citep{2014ApJS..214...15S} is smaller than ours, implying that strong [O {\footnotesize \uppercase\expandafter{\romannumeral3}}]$\lambda$5007 compact galaxies possibly have different mass assembly and evolution history. They assemble their mass faster. We also present the galaxies in the epoch of reionization (EoR) from \cite{2022ApJ...927..170T} in Figure \ref{fig:MSFR}. From this figure, we can see that most of EoR galaxies are located close to the main sequence of our samples at high-mass end, showing the high SFMS consistency. It suggests that our strong [O {\footnotesize \uppercase\expandafter{\romannumeral3}}]$\lambda$5007 compact galaxies are likely analogous to those high-redshift EoR galaxies.

\subsection{Mass-Metallicity Relation} \label{subsec:MZR}
\begin{figure}[htbp]
\centering
\includegraphics [width=\linewidth]{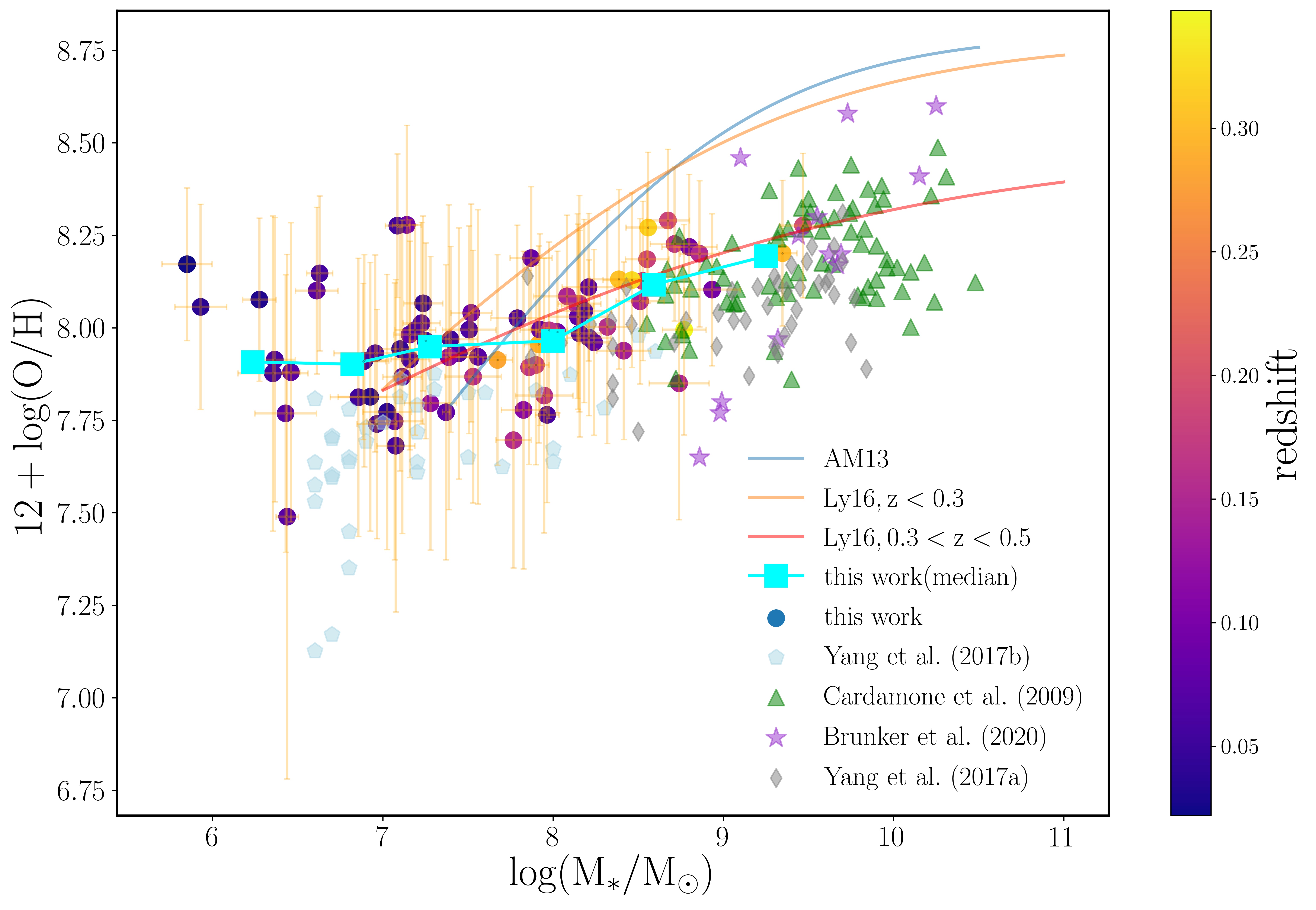}
\caption{Mass-metallicity relation of our strong [O {\footnotesize \uppercase\expandafter{\romannumeral3}}]$\lambda$5007 compact galaxies (circles) color-coded by redshift. The orange error bars represent the measurement uncertainties. The other points are the same as those in Figure \ref{fig:MSFR}. The cyan squares display the median metallities in different mass bins. The blue solid curve is the MZR from \cite{2013ApJ...765..140A}. The yellow and green solid curves are MZRs from \cite{2016ApJ...828...67L} in different redshift ranges.
\label{fig:MZR}}
\end{figure}

The mass-metallicity relation (MZR) suggests that more massive galaxies tend to be more metal-rich, and the trend holds from local to distant universe \citep[e.g.,][]{1979A&A....80..155L, 2004ApJ...613..898T, 2008ASPC..396..409M}. Figure \ref{fig:MZR} shows the mass-metallicity relation of our galaxy sample. The median metallicities in several mass bins are calculated, which are plotted in cyan. For comparisons, the local MZR from the stacked spectra by \cite{2013ApJ...765..140A} is presented in blue solid line, while the MZRs for star forming galaxies at ${z < 0.3}$ and $0.3 < {z} < 0.5$ derived by \cite{2016ApJ...828...67L} are shown in yellow and green solid lines, respectively. We also compare our MZRs with those of other strong [O {\footnotesize \uppercase\expandafter{\romannumeral3}}]$\lambda$5007 sample as mentioned before. In order for consistent comparison, the metallicity of green pea galaxies in \cite{2009MNRAS.399.1191C} is recalculated using the same N2-based method as used in this paper. \cite{2020ApJ...898...68B} uses the O3N2 method to calculate the gas-phase metallicity and \cite{2017ApJ...844..171Y} and \cite{2017ApJ...847...38Y} uses the direct $T_e$ method. As shown in Figure \ref{fig:MZR}, our compact galaxies together with other strong strong [O {\footnotesize \uppercase\expandafter{\romannumeral3}}]$\lambda$5007 sample present a much flatter MZR comparing to the local MZRs. Our MZR is similar to the relation in \cite{2022ApJ...927...57L}, and is close to the one at $0.3 < {z} < 0.5$ derived by \cite{2016ApJ...828...67L}. The flatter MZR indicates that strong [O {\footnotesize \uppercase\expandafter{\romannumeral3}}]$\lambda$5007 compact galaxies might have slower enrichment and are probably on the early stage of star formation.

\subsection{Environment} \label{environment}
\begin{figure}[htbp]
\centering
\includegraphics [width=\linewidth]{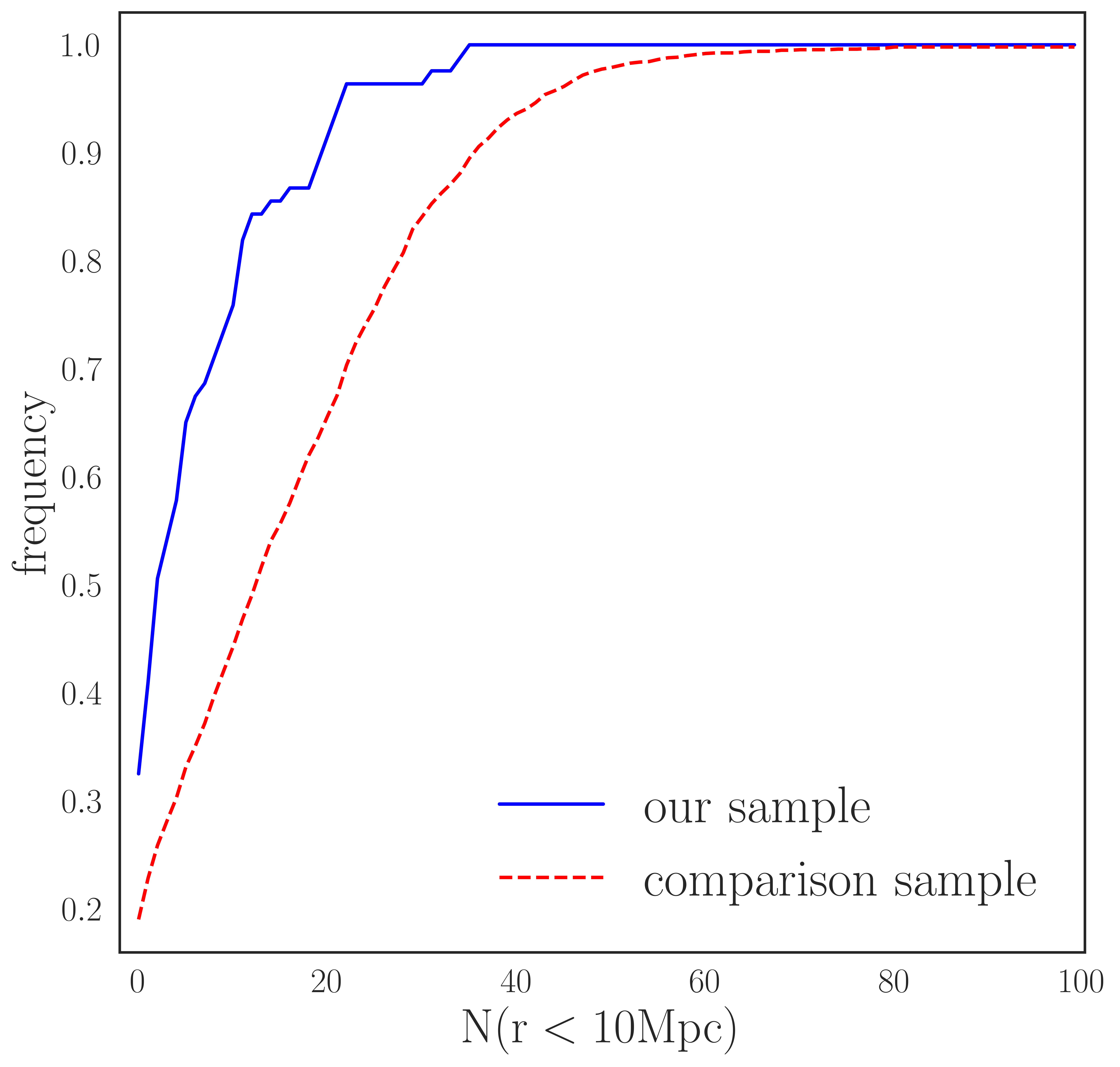}
\caption{Cumulative probability distribution of the local density for our galaxies (solid blue line) and that of a comparison sample dashed red line).  
\label{fig:env}}
\end{figure}

It was stated in \cite{2009MNRAS.399.1191C} that green pea galaxies reside in low-density environments. We calculate the local density for each galaxy in our sample in the 3D cosmological space, which is the number of neighbour galaxies within a distance of 10 Mpc. A control galaxy sample is randomly selected from eBOSS, including typical star-forming galaxies in the same redshift range of our galaxies. Figure \ref{fig:env} shows the cumulative probability distribution (CDF) of the local density for our galaxy sample and that of the control sample. The CDF of our sample lies above the one of the comparison sample, suggesting that strong [O {\footnotesize \uppercase\expandafter{\romannumeral3}}]$\lambda$5007 compact galaxies inhabit  low-density regions.


\section{SUMMARY} \label{sec:summary}

In this paper, we have assembled a new set of 76 strong [O {\footnotesize \uppercase\expandafter{\romannumeral3}}]$\lambda$5007 compact galaxies from SDSS DR16 that spans a redshift range of $0.02 < z < 0.35$. They share common properties with previously studied rare galaxies in the local universe: green pea and blueberry galaxies. All of these galaxies are compact and have extremely strong [O {\footnotesize \uppercase\expandafter{\romannumeral3}}]$\lambda$5007, but cover different redshift ranges, so they should have the same physical origin and evolution. We explore their detailed properties with the eBOSS spectra and multi-wavelength data, including scaling relations and environment etc. The main conclusion of this paper is as follows.
\begin{enumerate}
    \item[1)] Our strong [O {\footnotesize \uppercase\expandafter{\romannumeral3}}]$\lambda$5007 compact galaxies present relatively high SFRs (meidan $\rm{\sim 2.81 M_{\odot} yr^{-1}}$), low stellar mass (median log$\rm{M_*}$ $\sim 7.98$) and low metallicity (median $\rm{log[O/H] + 12 \sim 7.96}$). Their abnormally large equivalent widths of [O {\footnotesize \uppercase\expandafter{\romannumeral3}}]$\lambda$5007 make them sufficiently different from normal star-forming galaxies. 
    \item[2)] In all diagnostic diagrams including EW([O {\footnotesize \uppercase\expandafter{\romannumeral3}}]) vs. SFR, mass-metallicity relation, and main sequence, the galaxies at 0.112 $<$ z $<$ 0.36 in our sample have the same properties of green pea galaxies and those at lower redshift share common properties with blueberry galaxies. It indicates that such previous studied compact galaxies follow the same physical origin and evolutionary path.
    \item[3)] From the main sequence relation, we show that the SFR of our sample increases with redshift. For specified stellar mass, the SFR and corresponding sSFR are higher than those of normal star-forming galaxies. In addition, the slope of the main-sequence relation for our galaxies is about 0.69, higher than that of normal galaxies, indicating strong [O {\footnotesize \uppercase\expandafter{\romannumeral3}}]$\lambda$5007 compact galaxies assemble their stellar mass faster.
    \item[4)] The MZR diagram shows most of our galaxies are metal-poor. It has a much shallower slope than those of normal galaxies in local universe. Our MZR is close to the one  derived by \cite{2016ApJ...828...67L} at a higher redshift range. The flatter MZR implies that strong [O {\footnotesize \uppercase\expandafter{\romannumeral3}}]$\lambda$5007 compact galaxies have slower chemical enrichment relative to those normal galaxies at the same redshift and hence are on the earlier stage of the galaxy evolution. 
    \item[5)] The local density is calculated for each compact galaxy in our sample. In contrast to the control sample randomly selected from SDSS DR16, we confirm that our strong [O {\footnotesize \uppercase\expandafter{\romannumeral3}}]$\lambda$5007 compact galaxies reside in relatively low-density environment. 
\end{enumerate}

Ongoing large-scale spectroscopic surveys, such as Dark Energy Spectroscopic Instrument \citep[DESI;][]{2016arXiv161100036D}, will provide great opportunities for thoroughly understanding such compact galaxies with extremely strong characteristic emission lines.
\begin{acknowledgments} 
This work is supported by the National Natural Science Foundation of China (NSFC, Nos. 12120101003 and 11890691), National Key R\&D Program of China (Grant No. 2022YFA1602902), Beijing Municipal Natural Science Foundation (Grant No. 1222028). This work is also supported by the China Manned Space Project (Nos. CMS-CSST-2021-A02 and CMS-CSST-2021-A04). XK acknowledges the supports from the Strategic Priority Research Program of Chinese Academy of Sciences (No. XDB 41000000), NSFC (Nos.12233008 and 11973038), China Manned Space Project (Grant No. CMS-CSST-2021-A07), the Cyrus Chun Ying Tang Foundations and the Frontier Scientific Research Program of Deep Space Exploration Laboratory. 

\end{acknowledgments}
\bibliography{sample631}{}

\begin{thebibliography}{}
\expandafter\ifx\csname natexlab\endcsname\relax\def\natexlab#1{#1}\fi
\providecommand{\url}[1]{\href{#1}{#1}}
\providecommand{\dodoi}[1]{doi:~\href{http://doi.org/#1}{\nolinkurl{#1}}}
\providecommand{\doeprint}[1]{\href{http://ascl.net/#1}{\nolinkurl{http://ascl.net/#1}}}
\providecommand{\doarXiv}[1]{\href{https://arxiv.org/abs/#1}{\nolinkurl{https://arxiv.org/abs/#1}}}

\bibitem[{{Ahumada} {et~al.}(2020){Ahumada}, {Prieto}, {Almeida}, {Anders},
  {Anderson}, {Andrews}, {Anguiano}, {Arcodia}, {Armengaud}, {Aubert}, {Avila},
  {Avila-Reese}, {Badenes}, {Balland}, {Barger}, {Barrera-Ballesteros}, {Basu},
  {Bautista}, {Beaton}, {Beers}, {Benavides}, {Bender}, {Bernardi}, {Bershady},
  {Beutler}, {Bidin}, {Bird}, {Bizyaev}, {Blanc}, {Blanton}, {Boquien},
  {Borissova}, {Bovy}, {Brandt}, {Brinkmann}, {Brownstein}, {Bundy}, {Bureau},
  {Burgasser}, {Burtin}, {Cano-D{\'\i}az}, {Capasso}, {Cappellari}, {Carrera},
  {Chabanier}, {Chaplin}, {Chapman}, {Cherinka}, {Chiappini}, {Doohyun Choi},
  {Chojnowski}, {Chung}, {Clerc}, {Coffey}, {Comerford}, {Comparat}, {da
  Costa}, {Cousinou}, {Covey}, {Crane}, {Cunha}, {Ilha}, {Dai}, {Damsted},
  {Darling}, {Davidson}, {Davies}, {Dawson}, {De}, {de la Macorra}, {De Lee},
  {Queiroz}, {Deconto Machado}, {de la Torre}, {Dell'Agli}, {du Mas des
  Bourboux}, {Diamond-Stanic}, {Dillon}, {Donor}, {Drory}, {Duckworth},
  {Dwelly}, {Ebelke}, {Eftekharzadeh}, {Davis Eigenbrot}, {Elsworth},
  {Eracleous}, {Erfanianfar}, {Escoffier}, {Fan}, {Farr},
  {Fern{\'a}ndez-Trincado}, {Feuillet}, {Finoguenov}, {Fofie},
  {Fraser-McKelvie}, {Frinchaboy}, {Fromenteau}, {Fu}, {Galbany}, {Garcia},
  {Garc{\'\i}a-Hern{\'a}ndez}, {Oehmichen}, {Ge}, {Maia}, {Geisler}, {Gelfand},
  {Goddy}, {Gonzalez-Perez}, {Grabowski}, {Green}, {Grier}, {Guo}, {Guy},
  {Harding}, {Hasselquist}, {Hawken}, {Hayes}, {Hearty}, {Hekker}, {Hogg},
  {Holtzman}, {Horta}, {Hou}, {Hsieh}, {Huber}, {Hunt}, {Chitham}, {Imig},
  {Jaber}, {Angel}, {Johnson}, {Jones}, {J{\"o}nsson}, {Jullo}, {Kim},
  {Kinemuchi}, {Kirkpatrick}, {Kite}, {Klaene}, {Kneib}, {Kollmeier}, {Kong},
  {Kounkel}, {Krishnarao}, {Lacerna}, {Lan}, {Lane}, {Law}, {Le Goff}, {Leung},
  {Lewis}, {Li}, {Lian}, {Lin}, {Long}, {Longa-Pe{\~n}a}, {Lundgren}, {Lyke},
  {Ted Mackereth}, {MacLeod}, {Majewski}, {Manchado}, {Maraston}, {Martini},
  {Masseron}, {Masters}, {Mathur}, {McDermid}, {Merloni}, {Merrifield},
  {M{\'e}sz{\'a}ros}, {Miglio}, {Minniti}, {Minsley}, {Miyaji}, {Mohammad},
  {Mosser}, {Mueller}, {Muna}, {Mu{\~n}oz-Guti{\'e}rrez}, {Myers}, {Nadathur},
  {Nair}, {Nandra}, {do Nascimento}, {Nevin}, {Newman}, {Nidever}, {Nitschelm},
  {Noterdaeme}, {O'Connell}, {Olmstead}, {Oravetz}, {Oravetz}, {Osorio},
  {Pace}, {Padilla}, {Palanque-Delabrouille}, {Palicio}, {Pan}, {Pan},
  {Parker}, {Paviot}, {Peirani}, {Ram{\'r}ez}, {Penny}, {Percival},
  {Perez-Fournon}, {P{\'e}rez-R{\`a}fols}, {Petitjean}, {Pieri},
  {Pinsonneault}, {Poovelil}, {Povick}, {Prakash}, {Price-Whelan}, {Raddick},
  {Raichoor}, {Ray}, {Rembold}, {Rezaie}, {Riffel}, {Riffel}, {Rix}, {Robin},
  {Roman-Lopes}, {Rom{\'a}n-Z{\'u}{\~n}iga}, {Rose}, {Ross}, {Rossi},
  {Rowlands}, {Rubin}, {Salvato}, {S{\'a}nchez}, {S{\'a}nchez-Menguiano},
  {S{\'a}nchez-Gallego}, {Sayres}, {Schaefer}, {Schiavon}, {Schimoia},
  {Schlafly}, {Schlegel}, {Schneider}, {Schultheis}, {Schwope}, {Seo},
  {Serenelli}, {Shafieloo}, {Shamsi}, {Shao}, {Shen}, {Shetrone}, {Shirley},
  {Aguirre}, {Simon}, {Skrutskie}, {Slosar}, {Smethurst}, {Sobeck}, {Sodi},
  {Souto}, {Stark}, {Stassun}, {Steinmetz}, {Stello}, {Stermer},
  {Storchi-Bergmann}, {Streblyanska}, {Stringfellow}, {Stutz}, {Su{\'a}rez},
  {Sun}, {Taghizadeh-Popp}, {Talbot}, {Tayar}, {Thakar}, {Theriault}, {Thomas},
  {Thomas}, {Tinker}, {Tojeiro}, {Toledo}, {Tremonti}, {Troup}, {Tuttle},
  {Unda-Sanzana}, {Valentini}, {Vargas-Gonz{\'a}lez}, {Vargas-Maga{\~n}a},
  {V{\'a}zquez-Mata}, {Vivek}, {Wake}, {Wang}, {Weaver}, {Weijmans}, {Wild},
  {Wilson}, {Wilson}, {Wolthuis}, {Wood-Vasey}, {Yan}, {Yang}, {Y{\`e}che},
  {Zamora}, {Zarrouk}, {Zasowski}, {Zhang}, {Zhao}, {Zhao}, {Zheng}, {Zheng},
  {Zhu}, \& {Zou}}]{2020ApJS..249....3A}
{Ahumada}, R., {Prieto}, C.~A., {Almeida}, A., {et~al.} 2020, \apjs, 249, 3,
  \dodoi{10.3847/1538-4365/ab929e}

\bibitem[{{Aller}(1984)}]{1984ASSL..112.....A}
{Aller}, L.~H. 1984, {Physics of thermal gaseous nebulae},
  \dodoi{10.1007/978-94-010-9639-3}

\bibitem[{{Andrews} \& {Martini}(2013)}]{2013ApJ...765..140A}
{Andrews}, B.~H., \& {Martini}, P. 2013, \apj, 765, 140,
  \dodoi{10.1088/0004-637X/765/2/140}

\bibitem[{Baldwin {et~al.}(1981)Baldwin, Phillips, \& Terlevich}]{Baldwin_1981}
Baldwin, J.~A., Phillips, M.~M., \& Terlevich, R. 1981, Publications of the
  Astronomical Society of the Pacific, 93, 5, \dodoi{10.1086/130766}

\bibitem[{{Blanton} {et~al.}(2017){Blanton}, {Bershady}, {Abolfathi},
  {Albareti}, {Allende Prieto}, {Almeida}, {Alonso-Garc{\'\i}a}, {Anders},
  {Anderson}, {Andrews}, {Aquino-Ort{\'\i}z}, {Arag{\'o}n-Salamanca},
  {Argudo-Fern{\'a}ndez}, {Armengaud}, {Aubourg}, {Avila-Reese}, {Badenes},
  {Bailey}, {Barger}, {Barrera-Ballesteros}, {Bartosz}, {Bates}, {Baumgarten},
  {Bautista}, {Beaton}, {Beers}, {Belfiore}, {Bender}, {Berlind}, {Bernardi},
  {Beutler}, {Bird}, {Bizyaev}, {Blanc}, {Blomqvist}, {Bolton}, {Boquien},
  {Borissova}, {van den Bosch}, {Bovy}, {Brandt}, {Brinkmann}, {Brownstein},
  {Bundy}, {Burgasser}, {Burtin}, {Busca}, {Cappellari}, {Delgado Carigi},
  {Carlberg}, {Carnero Rosell}, {Carrera}, {Chanover}, {Cherinka}, {Cheung},
  {G{\'o}mez Maqueo Chew}, {Chiappini}, {Choi}, {Chojnowski}, {Chuang},
  {Chung}, {Cirolini}, {Clerc}, {Cohen}, {Comparat}, {da Costa}, {Cousinou},
  {Covey}, {Crane}, {Croft}, {Cruz-Gonzalez}, {Garrido Cuadra}, {Cunha},
  {Damke}, {Darling}, {Davies}, {Dawson}, {de la Macorra}, {Dell'Agli}, {De
  Lee}, {Delubac}, {Di Mille}, {Diamond-Stanic}, {Cano-D{\'\i}az}, {Donor},
  {Downes}, {Drory}, {du Mas des Bourboux}, {Duckworth}, {Dwelly}, {Dyer},
  {Ebelke}, {Eigenbrot}, {Eisenstein}, {Emsellem}, {Eracleous}, {Escoffier},
  {Evans}, {Fan}, {Fern{\'a}ndez-Alvar}, {Fernandez-Trincado}, {Feuillet},
  {Finoguenov}, {Fleming}, {Font-Ribera}, {Fredrickson}, {Freischlad},
  {Frinchaboy}, {Fuentes}, {Galbany}, {Garcia-Dias},
  {Garc{\'\i}a-Hern{\'a}ndez}, {Gaulme}, {Geisler}, {Gelfand},
  {Gil-Mar{\'\i}n}, {Gillespie}, {Goddard}, {Gonzalez-Perez}, {Grabowski},
  {Green}, {Grier}, {Gunn}, {Guo}, {Guy}, {Hagen}, {Hahn}, {Hall}, {Harding},
  {Hasselquist}, {Hawley}, {Hearty}, {Gonzalez Hern{\'a}ndez}, {Ho}, {Hogg},
  {Holley-Bockelmann}, {Holtzman}, {Holzer}, {Huehnerhoff}, {Hutchinson},
  {Hwang}, {Ibarra-Medel}, {da Silva Ilha}, {Ivans}, {Ivory}, {Jackson},
  {Jensen}, {Johnson}, {Jones}, {J{\"o}nsson}, {Jullo}, {Kamble}, {Kinemuchi},
  {Kirkby}, {Kitaura}, {Klaene}, {Knapp}, {Kneib}, {Kollmeier}, {Lacerna},
  {Lane}, {Lang}, {Law}, {Lazarz}, {Lee}, {Le Goff}, {Liang}, {Li}, {Li},
  {Lian}, {Lima}, {Lin}, {Lin}, {Bertran de Lis}, {Liu}, {de Icaza Lizaola},
  {Long}, {Lucatello}, {Lundgren}, {MacDonald}, {Deconto Machado}, {MacLeod},
  {Mahadevan}, {Geimba Maia}, {Maiolino}, {Majewski}, {Malanushenko},
  {Malanushenko}, {Manchado}, {Mao}, {Maraston}, {Marques-Chaves}, {Masseron},
  {Masters}, {McBride}, {McDermid}, {McGrath}, {McGreer}, {Medina Pe{\~n}a},
  {Melendez}, {Merloni}, {Merrifield}, {Meszaros}, {Meza}, {Minchev},
  {Minniti}, {Miyaji}, {More}, {Mulchaey}, {M{\"u}ller-S{\'a}nchez}, {Muna},
  {Munoz}, {Myers}, {Nair}, {Nandra}, {Correa do Nascimento}, {Negrete},
  {Ness}, {Newman}, {Nichol}, {Nidever}, {Nitschelm}, {Ntelis}, {O'Connell},
  {Oelkers}, {Oravetz}, {Oravetz}, {Pace}, {Padilla}, {Palanque-Delabrouille},
  {Alonso Palicio}, {Pan}, {Parejko}, {Parikh}, {P{\^a}ris}, {Park}, {Patten},
  {Peirani}, {Pellejero-Ibanez}, {Penny}, {Percival}, {Perez-Fournon},
  {Petitjean}, {Pieri}, {Pinsonneault}, {Pisani}, {Poleski}, {Prada},
  {Prakash}, {Queiroz}, {Raddick}, {Raichoor}, {Barboza Rembold}, {Richstein},
  {Riffel}, {Riffel}, {Rix}, {Robin}, {Rockosi}, {Rodr{\'\i}guez-Torres},
  {Roman-Lopes}, {Rom{\'a}n-Z{\'u}{\~n}iga}, {Rosado}, {Ross}, {Rossi}, {Ruan},
  {Ruggeri}, {Rykoff}, {Salazar-Albornoz}, {Salvato}, {S{\'a}nchez}, {Aguado},
  {S{\'a}nchez-Gallego}, {Santana}, {Santiago}, {Sayres}, {Schiavon}, {da Silva
  Schimoia}, {Schlafly}, {Schlegel}, {Schneider}, {Schultheis}, {Schuster},
  {Schwope}, {Seo}, {Shao}, {Shen}, {Shetrone}, {Shull}, {Simon}, {Skinner},
  {Skrutskie}, {Slosar}, {Smith}, {Sobeck}, {Sobreira}, {Somers}, {Souto},
  {Stark}, {Stassun}, {Stauffer}, {Steinmetz}, {Storchi-Bergmann},
  {Streblyanska}, {Stringfellow}, {Su{\'a}rez}, {Sun}, {Suzuki}, {Szigeti},
  {Taghizadeh-Popp}, {Tang}, {Tao}, {Tayar}, {Tembe}, {Teske}, {Thakar},
  {Thomas}, {Thompson}, {Tinker}, {Tissera}, {Tojeiro}, {Hernandez Toledo}, {de
  la Torre}, {Tremonti}, {Troup}, {Valenzuela}, {Martinez Valpuesta},
  {Vargas-Gonz{\'a}lez}, {Vargas-Maga{\~n}a}, {Vazquez}, {Villanova}, {Vivek},
  {Vogt}, {Wake}, {Walterbos}, {Wang}, {Weaver}, {Weijmans}, {Weinberg},
  {Westfall}, {Whelan}, {Wild}, {Wilson}, {Wood-Vasey}, {Wylezalek}, {Xiao},
  {Yan}, {Yang}, {Ybarra}, {Y{\`e}che}, {Zakamska}, {Zamora}, {Zarrouk},
  {Zasowski}, {Zhang}, {Zhao}, {Zheng}, {Zheng}, {Zhou}, {Zhou}, {Zhu},
  {Zoccali}, \& {Zou}}]{2017AJ....154...28B}
{Blanton}, M.~R., {Bershady}, M.~A., {Abolfathi}, B., {et~al.} 2017, \aj, 154,
  28, \dodoi{10.3847/1538-3881/aa7567}

\bibitem[{{Boquien} {et~al.}(2019){Boquien}, {Burgarella}, {Roehlly}, {Buat},
  {Ciesla}, {Corre}, {Inoue}, \& {Salas}}]{2019A&A...622A.103B}
{Boquien}, M., {Burgarella}, D., {Roehlly}, Y., {et~al.} 2019, \aap, 622, A103,
  \dodoi{10.1051/0004-6361/201834156}

\bibitem[{{Brunker} {et~al.}(2020){Brunker}, {Salzer}, {Janowiecki}, {Finn}, \&
  {Helou}}]{2020ApJ...898...68B}
{Brunker}, S.~W., {Salzer}, J.~J., {Janowiecki}, S., {Finn}, R.~A., \& {Helou},
  G. 2020, \apj, 898, 68, \dodoi{10.3847/1538-4357/ab9ec0}

\bibitem[{{Bruzual} \& {Charlot}(2003)}]{2003MNRAS.344.1000B}
{Bruzual}, G., \& {Charlot}, S. 2003, \mnras, 344, 1000,
  \dodoi{10.1046/j.1365-8711.2003.06897.x}

\bibitem[{{Burgarella} {et~al.}(2005){Burgarella}, {Buat}, \&
  {Iglesias-P{\'a}ramo}}]{2005MNRAS.360.1413B}
{Burgarella}, D., {Buat}, V., \& {Iglesias-P{\'a}ramo}, J. 2005, \mnras, 360,
  1413, \dodoi{10.1111/j.1365-2966.2005.09131.x}

\bibitem[{{Calzetti} {et~al.}(2000){Calzetti}, {Armus}, {Bohlin}, {Kinney},
  {Koornneef}, \& {Storchi-Bergmann}}]{2000ApJ...533..682C}
{Calzetti}, D., {Armus}, L., {Bohlin}, R.~C., {et~al.} 2000, \apj, 533, 682,
  \dodoi{10.1086/308692}

\bibitem[{{Cardamone} {et~al.}(2009){Cardamone}, {Schawinski}, {Sarzi},
  {Bamford}, {Bennert}, {Urry}, {Lintott}, {Keel}, {Parejko}, {Nichol},
  {Thomas}, {Andreescu}, {Murray}, {Raddick}, {Slosar}, {Szalay}, \&
  {Vandenberg}}]{2009MNRAS.399.1191C}
{Cardamone}, C., {Schawinski}, K., {Sarzi}, M., {et~al.} 2009, \mnras, 399,
  1191, \dodoi{10.1111/j.1365-2966.2009.15383.x}

\bibitem[{{Cid Fernandes} {et~al.}(2005){Cid Fernandes}, {Mateus}, {Sodr{\'e}},
  {Stasi{\'n}ska}, \& {Gomes}}]{2005MNRAS.358..363C}
{Cid Fernandes}, R., {Mateus}, A., {Sodr{\'e}}, L., {Stasi{\'n}ska}, G., \&
  {Gomes}, J.~M. 2005, \mnras, 358, 363,
  \dodoi{10.1111/j.1365-2966.2005.08752.x}

\bibitem[{{Dawson} {et~al.}(2013){Dawson}, {Schlegel}, {Ahn}, {Anderson},
  {Aubourg}, {Bailey}, {Barkhouser}, {Bautista}, {Beifiori}, {Berlind},
  {Bhardwaj}, {Bizyaev}, {Blake}, {Blanton}, {Blomqvist}, {Bolton}, {Borde},
  {Bovy}, {Brandt}, {Brewington}, {Brinkmann}, {Brown}, {Brownstein}, {Bundy},
  {Busca}, {Carithers}, {Carnero}, {Carr}, {Chen}, {Comparat}, {Connolly},
  {Cope}, {Croft}, {Cuesta}, {da Costa}, {Davenport}, {Delubac}, {de Putter},
  {Dhital}, {Ealet}, {Ebelke}, {Eisenstein}, {Escoffier}, {Fan}, {Filiz Ak},
  {Finley}, {Font-Ribera}, {G{\'e}nova-Santos}, {Gunn}, {Guo}, {Haggard},
  {Hall}, {Hamilton}, {Harris}, {Harris}, {Ho}, {Hogg}, {Holder}, {Honscheid},
  {Huehnerhoff}, {Jordan}, {Jordan}, {Kauffmann}, {Kazin}, {Kirkby}, {Klaene},
  {Kneib}, {Le Goff}, {Lee}, {Long}, {Loomis}, {Lundgren}, {Lupton}, {Maia},
  {Makler}, {Malanushenko}, {Malanushenko}, {Mandelbaum}, {Manera}, {Maraston},
  {Margala}, {Masters}, {McBride}, {McDonald}, {McGreer}, {McMahon}, {Mena},
  {Miralda-Escud{\'e}}, {Montero-Dorta}, {Montesano}, {Muna}, {Myers},
  {Naugle}, {Nichol}, {Noterdaeme}, {Nuza}, {Olmstead}, {Oravetz}, {Oravetz},
  {Owen}, {Padmanabhan}, {Palanque-Delabrouille}, {Pan}, {Parejko},
  {P{\^a}ris}, {Percival}, {P{\'e}rez-Fournon}, {P{\'e}rez-R{\`a}fols},
  {Petitjean}, {Pfaffenberger}, {Pforr}, {Pieri}, {Prada}, {Price-Whelan},
  {Raddick}, {Rebolo}, {Rich}, {Richards}, {Rockosi}, {Roe}, {Ross}, {Ross},
  {Rossi}, {Rubi{\~n}o-Martin}, {Samushia}, {S{\'a}nchez}, {Sayres}, {Schmidt},
  {Schneider}, {Sc{\'o}ccola}, {Seo}, {Shelden}, {Sheldon}, {Shen}, {Shu},
  {Slosar}, {Smee}, {Snedden}, {Stauffer}, {Steele}, {Strauss}, {Streblyanska},
  {Suzuki}, {Swanson}, {Tal}, {Tanaka}, {Thomas}, {Tinker}, {Tojeiro},
  {Tremonti}, {Vargas Maga{\~n}a}, {Verde}, {Viel}, {Wake}, {Watson}, {Weaver},
  {Weinberg}, {Weiner}, {West}, {White}, {Wood-Vasey}, {Yeche}, {Zehavi},
  {Zhao}, \& {Zheng}}]{2013AJ....145...10D}
{Dawson}, K.~S., {Schlegel}, D.~J., {Ahn}, C.~P., {et~al.} 2013, \aj, 145, 10,
  \dodoi{10.1088/0004-6256/145/1/10}

\bibitem[{{Dawson} {et~al.}(2016){Dawson}, {Kneib}, {Percival}, {Alam},
  {Albareti}, {Anderson}, {Armengaud}, {Aubourg}, {Bailey}, {Bautista},
  {Berlind}, {Bershady}, {Beutler}, {Bizyaev}, {Blanton}, {Blomqvist},
  {Bolton}, {Bovy}, {Brandt}, {Brinkmann}, {Brownstein}, {Burtin}, {Busca},
  {Cai}, {Chuang}, {Clerc}, {Comparat}, {Cope}, {Croft}, {Cruz-Gonzalez}, {da
  Costa}, {Cousinou}, {Darling}, {de la Macorra}, {de la Torre}, {Delubac}, {du
  Mas des Bourboux}, {Dwelly}, {Ealet}, {Eisenstein}, {Eracleous}, {Escoffier},
  {Fan}, {Finoguenov}, {Font-Ribera}, {Frinchaboy}, {Gaulme}, {Georgakakis},
  {Green}, {Guo}, {Guy}, {Ho}, {Holder}, {Huehnerhoff}, {Hutchinson}, {Jing},
  {Jullo}, {Kamble}, {Kinemuchi}, {Kirkby}, {Kitaura}, {Klaene}, {Laher},
  {Lang}, {Laurent}, {Le Goff}, {Li}, {Liang}, {Lima}, {Lin}, {Lin}, {Lin},
  {Long}, {Lundgren}, {MacDonald}, {Geimba Maia}, {Malanushenko},
  {Malanushenko}, {Mariappan}, {McBride}, {McGreer}, {M{\'e}nard}, {Merloni},
  {Meza}, {Montero-Dorta}, {Muna}, {Myers}, {Nandra}, {Naugle}, {Newman},
  {Noterdaeme}, {Nugent}, {Ogando}, {Olmstead}, {Oravetz}, {Oravetz},
  {Padmanabhan}, {Palanque-Delabrouille}, {Pan}, {Parejko}, {P{\^a}ris},
  {Peacock}, {Petitjean}, {Pieri}, {Pisani}, {Prada}, {Prakash}, {Raichoor},
  {Reid}, {Rich}, {Ridl}, {Rodriguez-Torres}, {Carnero Rosell}, {Ross},
  {Rossi}, {Ruan}, {Salvato}, {Sayres}, {Schneider}, {Schlegel}, {Seljak},
  {Seo}, {Sesar}, {Shandera}, {Shu}, {Slosar}, {Sobreira}, {Streblyanska},
  {Suzuki}, {Taylor}, {Tao}, {Tinker}, {Tojeiro}, {Vargas-Maga{\~n}a}, {Wang},
  {Weaver}, {Weinberg}, {White}, {Wood-Vasey}, {Yeche}, {Zhai}, {Zhao}, {Zhao},
  {Zheng}, {Ben Zhu}, \& {Zou}}]{2016AJ....151...44D}
{Dawson}, K.~S., {Kneib}, J.-P., {Percival}, W.~J., {et~al.} 2016, \aj, 151,
  44, \dodoi{10.3847/0004-6256/151/2/44}

\bibitem[{{DESI Collaboration} {et~al.}(2016){DESI Collaboration}, {Aghamousa},
  {Aguilar}, {Ahlen}, {Alam}, {Allen}, {Allende Prieto}, {Annis}, {Bailey},
  {Balland}, {Ballester}, {Baltay}, {Beaufore}, {Bebek}, {Beers}, {Bell},
  {Bernal}, {Besuner}, {Beutler}, {Blake}, {Bleuler}, {Blomqvist}, {Blum},
  {Bolton}, {Briceno}, {Brooks}, {Brownstein}, {Buckley-Geer}, {Burden},
  {Burtin}, {Busca}, {Cahn}, {Cai}, {Cardiel-Sas}, {Carlberg}, {Carton},
  {Casas}, {Castander}, {Cervantes-Cota}, {Claybaugh}, {Close}, {Coker},
  {Cole}, {Comparat}, {Cooper}, {Cousinou}, {Crocce}, {Cuby}, {Cunningham},
  {Davis}, {Dawson}, {de la Macorra}, {De Vicente}, {Delubac}, {Derwent},
  {Dey}, {Dhungana}, {Ding}, {Doel}, {Duan}, {Ealet}, {Edelstein},
  {Eftekharzadeh}, {Eisenstein}, {Elliott}, {Escoffier}, {Evatt}, {Fagrelius},
  {Fan}, {Fanning}, {Farahi}, {Farihi}, {Favole}, {Feng}, {Fernandez},
  {Findlay}, {Finkbeiner}, {Fitzpatrick}, {Flaugher}, {Flender}, {Font-Ribera},
  {Forero-Romero}, {Fosalba}, {Frenk}, {Fumagalli}, {Gaensicke}, {Gallo},
  {Garcia-Bellido}, {Gaztanaga}, {Pietro Gentile Fusillo}, {Gerard},
  {Gershkovich}, {Giannantonio}, {Gillet}, {Gonzalez-de-Rivera},
  {Gonzalez-Perez}, {Gott}, {Graur}, {Gutierrez}, {Guy}, {Habib}, {Heetderks},
  {Heetderks}, {Heitmann}, {Hellwing}, {Herrera}, {Ho}, {Holland}, {Honscheid},
  {Huff}, {Hutchinson}, {Huterer}, {Hwang}, {Illa Laguna}, {Ishikawa},
  {Jacobs}, {Jeffrey}, {Jelinsky}, {Jennings}, {Jiang}, {Jimenez}, {Johnson},
  {Joyce}, {Jullo}, {Juneau}, {Kama}, {Karcher}, {Karkar}, {Kehoe}, {Kennamer},
  {Kent}, {Kilbinger}, {Kim}, {Kirkby}, {Kisner}, {Kitanidis}, {Kneib},
  {Koposov}, {Kovacs}, {Koyama}, {Kremin}, {Kron}, {Kronig}, {Kueter-Young},
  {Lacey}, {Lafever}, {Lahav}, {Lambert}, {Lampton}, {Landriau}, {Lang},
  {Lauer}, {Le Goff}, {Le Guillou}, {Le Van Suu}, {Lee}, {Lee}, {Leitner},
  {Lesser}, {Levi}, {L'Huillier}, {Li}, {Liang}, {Lin}, {Linder}, {Loebman},
  {Luki{\'c}}, {Ma}, {MacCrann}, {Magneville}, {Makarem}, {Manera}, {Manser},
  {Marshall}, {Martini}, {Massey}, {Matheson}, {McCauley}, {McDonald},
  {McGreer}, {Meisner}, {Metcalfe}, {Miller}, {Miquel}, {Moustakas}, {Myers},
  {Naik}, {Newman}, {Nichol}, {Nicola}, {Nicolati da Costa}, {Nie}, {Niz},
  {Norberg}, {Nord}, {Norman}, {Nugent}, {O'Brien}, {Oh}, {Olsen}, {Padilla},
  {Padmanabhan}, {Padmanabhan}, {Palanque-Delabrouille}, {Palmese},
  {Pappalardo}, {P{\^a}ris}, {Park}, {Patej}, {Peacock}, {Peiris}, {Peng},
  {Percival}, {Perruchot}, {Pieri}, {Pogge}, {Pollack}, {Poppett}, {Prada},
  {Prakash}, {Probst}, {Rabinowitz}, {Raichoor}, {Ree}, {Refregier}, {Regal},
  {Reid}, {Reil}, {Rezaie}, {Rockosi}, {Roe}, {Ronayette}, {Roodman}, {Ross},
  {Ross}, {Rossi}, {Rozo}, {Ruhlmann-Kleider}, {Rykoff}, {Sabiu}, {Samushia},
  {Sanchez}, {Sanchez}, {Schlegel}, {Schneider}, {Schubnell}, {Secroun},
  {Seljak}, {Seo}, {Serrano}, {Shafieloo}, {Shan}, {Sharples}, {Sholl},
  {Shourt}, {Silber}, {Silva}, {Sirk}, {Slosar}, {Smith}, {Smoot}, {Som},
  {Song}, {Sprayberry}, {Staten}, {Stefanik}, {Tarle}, {Sien Tie}, {Tinker},
  {Tojeiro}, {Valdes}, {Valenzuela}, {Valluri}, {Vargas-Magana}, {Verde},
  {Walker}, {Wang}, {Wang}, {Weaver}, {Weaverdyck}, {Wechsler}, {Weinberg},
  {White}, {Yang}, {Yeche}, {Zhang}, {Zhao}, {Zheng}, {Zhou}, {Zhou}, {Zhu},
  {Zou}, \& {Zu}}]{2016arXiv161100036D}
{DESI Collaboration}, {Aghamousa}, A., {Aguilar}, J., {et~al.} 2016, arXiv
  e-prints, arXiv:1611.00036, \dodoi{10.48550/arXiv.1611.00036}

\bibitem[{{Draine} {et~al.}(2014){Draine}, {Aniano}, {Krause}, {Groves},
  {Sandstrom}, {Braun}, {Leroy}, {Klaas}, {Linz}, {Rix}, {Schinnerer},
  {Schmiedeke}, \& {Walter}}]{2014ApJ...780..172D}
{Draine}, B.~T., {Aniano}, G., {Krause}, O., {et~al.} 2014, \apj, 780, 172,
  \dodoi{10.1088/0004-637X/780/2/172}

\bibitem[{{Fan} {et~al.}(2006){Fan}, {Strauss}, {Becker}, {White}, {Gunn},
  {Knapp}, {Richards}, {Schneider}, {Brinkmann}, \&
  {Fukugita}}]{2006AJ....132..117F}
{Fan}, X., {Strauss}, M.~A., {Becker}, R.~H., {et~al.} 2006, \aj, 132, 117,
  \dodoi{10.1086/504836}

\bibitem[{{Finkelstein} \& {Bagley}(2022)}]{2022ApJ...938...25F}
{Finkelstein}, S.~L., \& {Bagley}, M.~B. 2022, \apj, 938, 25,
  \dodoi{10.3847/1538-4357/ac89eb}

\bibitem[{{Finkelstein} {et~al.}(2019){Finkelstein}, {D'Aloisio},
  {Paardekooper}, {Ryan}, {Behroozi}, {Finlator}, {Livermore}, {Upton
  Sanderbeck}, {Dalla Vecchia}, \& {Khochfar}}]{2019ApJ...879...36F}
{Finkelstein}, S.~L., {D'Aloisio}, A., {Paardekooper}, J.-P., {et~al.} 2019,
  \apj, 879, 36, \dodoi{10.3847/1538-4357/ab1ea8}

\bibitem[{{Gao} {et~al.}(2018){Gao}, {Bao}, {Yuan}, {Kong}, {Zou}, {Zhou},
  {Gu}, {Lin}, {Liang}, \& {Huang}}]{2018ApJ...869...15G}
{Gao}, Y., {Bao}, M., {Yuan}, Q., {et~al.} 2018, \apj, 869, 15,
  \dodoi{10.3847/1538-4357/aae9ef}

\bibitem[{{Gunn} {et~al.}(2006){Gunn}, {Siegmund}, {Mannery}, {Owen}, {Hull},
  {Leger}, {Carey}, {Knapp}, {York}, {Boroski}, {Kent}, {Lupton}, {Rockosi},
  {Evans}, {Waddell}, {Anderson}, {Annis}, {Barentine}, {Bartoszek}, {Bastian},
  {Bracker}, {Brewington}, {Briegel}, {Brinkmann}, {Brown}, {Carr},
  {Czarapata}, {Drennan}, {Dombeck}, {Federwitz}, {Gillespie}, {Gonzales},
  {Hansen}, {Harvanek}, {Hayes}, {Jordan}, {Kinney}, {Klaene}, {Kleinman},
  {Kron}, {Kresinski}, {Lee}, {Limmongkol}, {Lindenmeyer}, {Long}, {Loomis},
  {McGehee}, {Mantsch}, {Neilsen}, {Neswold}, {Newman}, {Nitta}, {Peoples},
  {Pier}, {Prieto}, {Prosapio}, {Rivetta}, {Schneider}, {Snedden}, \&
  {Wang}}]{2006AJ....131.2332G}
{Gunn}, J.~E., {Siegmund}, W.~A., {Mannery}, E.~J., {et~al.} 2006, \aj, 131,
  2332, \dodoi{10.1086/500975}

\bibitem[{{Haiman} \& {Loeb}(1998)}]{1998ApJ...503..505H}
{Haiman}, Z., \& {Loeb}, A. 1998, \apj, 503, 505, \dodoi{10.1086/306017}

\bibitem[{{Henry} {et~al.}(2015){Henry}, {Scarlata}, {Martin}, \&
  {Erb}}]{2015ApJ...809...19H}
{Henry}, A., {Scarlata}, C., {Martin}, C.~L., \& {Erb}, D. 2015, \apj, 809, 19,
  \dodoi{10.1088/0004-637X/809/1/19}

\bibitem[{{Izotov} {et~al.}(2011){Izotov}, {Guseva}, \&
  {Thuan}}]{2011ApJ...728..161I}
{Izotov}, Y.~I., {Guseva}, N.~G., \& {Thuan}, T.~X. 2011, \apj, 728, 161,
  \dodoi{10.1088/0004-637X/728/2/161}

\bibitem[{{Izotov} {et~al.}(2020){Izotov}, {Schaerer}, {Worseck}, {Verhamme},
  {Guseva}, {Thuan}, {Orlitov{\'a}}, \& {Fricke}}]{2020MNRAS.491..468I}
{Izotov}, Y.~I., {Schaerer}, D., {Worseck}, G., {et~al.} 2020, \mnras, 491,
  468, \dodoi{10.1093/mnras/stz3041}

\bibitem[{{Jiang} {et~al.}(2019){Jiang}, {Malhotra}, {Rhoads}, \&
  {Yang}}]{2019ApJ...872..145J}
{Jiang}, T., {Malhotra}, S., {Rhoads}, J.~E., \& {Yang}, H. 2019, \apj, 872,
  145, \dodoi{10.3847/1538-4357/aaee8a}

\bibitem[{{Kauffmann} {et~al.}(2003){Kauffmann}, {Heckman}, {Tremonti},
  {Brinchmann}, {Charlot}, {White}, {Ridgway}, {Brinkmann}, {Fukugita}, {Hall},
  {Ivezi{\'c}}, {Richards}, \& {Schneider}}]{2003MNRAS.346.1055K}
{Kauffmann}, G., {Heckman}, T.~M., {Tremonti}, C., {et~al.} 2003, \mnras, 346,
  1055, \dodoi{10.1111/j.1365-2966.2003.07154.x}

\bibitem[{{Kennicutt}(1998)}]{1998ARA&A..36..189K}
{Kennicutt}, Robert~C., J. 1998, \araa, 36, 189,
  \dodoi{10.1146/annurev.astro.36.1.189}

\bibitem[{{Kewley} {et~al.}(2001){Kewley}, {Dopita}, {Sutherland}, {Heisler},
  \& {Trevena}}]{2001ApJ...556..121K}
{Kewley}, L.~J., {Dopita}, M.~A., {Sutherland}, R.~S., {Heisler}, C.~A., \&
  {Trevena}, J. 2001, \apj, 556, 121, \dodoi{10.1086/321545}

\bibitem[{{Kobulnicky} \& {Kewley}(2004)}]{2004ApJ...617..240K}
{Kobulnicky}, H.~A., \& {Kewley}, L.~J. 2004, \apj, 617, 240,
  \dodoi{10.1086/425299}

\bibitem[{{Lequeux} {et~al.}(1979){Lequeux}, {Peimbert}, {Rayo}, {Serrano}, \&
  {Torres-Peimbert}}]{1979A&A....80..155L}
{Lequeux}, J., {Peimbert}, M., {Rayo}, J.~F., {Serrano}, A., \&
  {Torres-Peimbert}, S. 1979, \aap, 80, 155

\bibitem[{{Lintott} {et~al.}(2008){Lintott}, {Schawinski}, {Slosar}, {Land},
  {Bamford}, {Thomas}, {Raddick}, {Nichol}, {Szalay}, {Andreescu}, {Murray}, \&
  {Vandenberg}}]{2008MNRAS.389.1179L}
{Lintott}, C.~J., {Schawinski}, K., {Slosar}, A., {et~al.} 2008, \mnras, 389,
  1179, \dodoi{10.1111/j.1365-2966.2008.13689.x}

\bibitem[{{Liu} {et~al.}(2022){Liu}, {Luo}, {Yang}, {Shen}, {Wang}, {Zhang},
  {Zheng}, {Song}, {Kong}, {Wang}, \& {Chen}}]{2022ApJ...927...57L}
{Liu}, S., {Luo}, A.~L., {Yang}, H., {et~al.} 2022, \apj, 927, 57,
  \dodoi{10.3847/1538-4357/ac4bd9}

\bibitem[{{Ly} {et~al.}(2014){Ly}, {Malkan}, {Nagao}, {Kashikawa}, {Shimasaku},
  \& {Hayashi}}]{2014ApJ...780..122L}
{Ly}, C., {Malkan}, M.~A., {Nagao}, T., {et~al.} 2014, \apj, 780, 122,
  \dodoi{10.1088/0004-637X/780/2/122}

\bibitem[{{Ly} {et~al.}(2016){Ly}, {Malkan}, {Rigby}, \&
  {Nagao}}]{2016ApJ...828...67L}
{Ly}, C., {Malkan}, M.~A., {Rigby}, J.~R., \& {Nagao}, T. 2016, \apj, 828, 67,
  \dodoi{10.3847/0004-637X/828/2/67}

\bibitem[{{Madau} {et~al.}(2004){Madau}, {Rees}, {Volonteri}, {Haardt}, \&
  {Oh}}]{2004ApJ...604..484M}
{Madau}, P., {Rees}, M.~J., {Volonteri}, M., {Haardt}, F., \& {Oh}, S.~P. 2004,
  \apj, 604, 484, \dodoi{10.1086/381935}

\bibitem[{{Maiolino} {et~al.}(2008{\natexlab{a}}){Maiolino}, {Nagao},
  {Grazian}, {Cocchia}, {Marconi}, {Mannucci}, {Cimatti}, {Pipino}, {Ballero},
  {Calura}, {Chiappini}, {Fontana}, {Granato}, {Matteucci}, {Pastorini},
  {Pentericci}, {Risaliti}, {Salvati}, \& {Silva}}]{2008A&A...488..463M}
{Maiolino}, R., {Nagao}, T., {Grazian}, A., {et~al.} 2008{\natexlab{a}}, \aap,
  488, 463, \dodoi{10.1051/0004-6361:200809678}

\bibitem[{{Maiolino} {et~al.}(2008{\natexlab{b}}){Maiolino}, {Nagao},
  {Grazian}, {Cocchia}, {Marconi}, {Mannucci}, {Cimatti}, {Pipino}, {Fontana},
  {Granato}, {Matteucci}, {Pentericci}, {Risaliti}, {Salvati}, \&
  {Silva}}]{2008ASPC..396..409M}
{Maiolino}, R., {Nagao}, T., {Grazian}, A., {et~al.} 2008{\natexlab{b}}, in
  Astronomical Society of the Pacific Conference Series, Vol. 396, Formation
  and Evolution of Galaxy Disks, ed. J.~G. {Funes} \& E.~M. {Corsini}, 409

\bibitem[{{Malkan} {et~al.}(2021){Malkan}, {Scully}, \&
  {Stecker}}]{2021ApJ...909...52M}
{Malkan}, M.~A., {Scully}, S.~T., \& {Stecker}, F.~W. 2021, \apj, 909, 52,
  \dodoi{10.3847/1538-4357/abda3f}

\bibitem[{{Markwardt}(2009)}]{2009ASPC..411..251M}
{Markwardt}, C.~B. 2009, in Astronomical Society of the Pacific Conference
  Series, Vol. 411, Astronomical Data Analysis Software and Systems XVIII, ed.
  D.~A. {Bohlender}, D.~{Durand}, \& P.~{Dowler}, 251.
\newblock \doarXiv{0902.2850}

\bibitem[{{McGreer} {et~al.}(2015){McGreer}, {Mesinger}, \&
  {D'Odorico}}]{2015MNRAS.447..499M}
{McGreer}, I.~D., {Mesinger}, A., \& {D'Odorico}, V. 2015, \mnras, 447, 499,
  \dodoi{10.1093/mnras/stu2449}

\bibitem[{{Nakajima} \& {Ouchi}(2014)}]{2014MNRAS.442..900N}
{Nakajima}, K., \& {Ouchi}, M. 2014, \mnras, 442, 900,
  \dodoi{10.1093/mnras/stu902}

\bibitem[{{Noll} {et~al.}(2009){Noll}, {Burgarella}, {Giovannoli}, {Buat},
  {Marcillac}, \& {Mu{\~n}oz-Mateos}}]{2009A&A...507.1793N}
{Noll}, S., {Burgarella}, D., {Giovannoli}, E., {et~al.} 2009, \aap, 507, 1793,
  \dodoi{10.1051/0004-6361/200912497}

\bibitem[{{Pettini} \& {Pagel}(2004)}]{2004MNRAS.348L..59P}
{Pettini}, M., \& {Pagel}, B. E.~J. 2004, \mnras, 348, L59,
  \dodoi{10.1111/j.1365-2966.2004.07591.x}

\bibitem[{{Rivera-Thorsen} {et~al.}(2015){Rivera-Thorsen}, {Hayes},
  {{\"O}stlin}, {Duval}, {Orlitov{\'a}}, {Verhamme}, {Mas-Hesse}, {Schaerer},
  {Cannon}, {Ot{\'\i}-Floranes}, {Sandberg}, {Guaita}, {Adamo}, {Atek},
  {Herenz}, {Kunth}, {Laursen}, \& {Melinder}}]{2015ApJ...805...14R}
{Rivera-Thorsen}, T.~E., {Hayes}, M., {{\"O}stlin}, G., {et~al.} 2015, \apj,
  805, 14, \dodoi{10.1088/0004-637X/805/1/14}

\bibitem[{{Robertson} {et~al.}(2010){Robertson}, {Ellis}, {Dunlop}, {McLure},
  \& {Stark}}]{2010Natur.468...49R}
{Robertson}, B.~E., {Ellis}, R.~S., {Dunlop}, J.~S., {McLure}, R.~J., \&
  {Stark}, D.~P. 2010, \nat, 468, 49, \dodoi{10.1038/nature09527}

\bibitem[{{Salpeter}(1955)}]{1955ApJ...121..161S}
{Salpeter}, E.~E. 1955, \apj, 121, 161, \dodoi{10.1086/145971}

\bibitem[{{Salzer} {et~al.}(2000){Salzer}, {Gronwall}, {Sarajedini}, \&
  {Chomiuk}}]{2000AAS...197.7606S}
{Salzer}, J.~J., {Gronwall}, C., {Sarajedini}, V.~L., \& {Chomiuk}, L.~B. 2000,
  in American Astronomical Society Meeting Abstracts, Vol. 197, American
  Astronomical Society Meeting Abstracts, 76.06

\bibitem[{{Speagle} {et~al.}(2014){Speagle}, {Steinhardt}, {Capak}, \&
  {Silverman}}]{2014ApJS..214...15S}
{Speagle}, J.~S., {Steinhardt}, C.~L., {Capak}, P.~L., \& {Silverman}, J.~D.
  2014, \apjs, 214, 15, \dodoi{10.1088/0067-0049/214/2/15}

\bibitem[{{Tacchella} {et~al.}(2022){Tacchella}, {Finkelstein}, {Bagley},
  {Dickinson}, {Ferguson}, {Giavalisco}, {Graziani}, {Grogin}, {Hathi},
  {Hutchison}, {Jung}, {Koekemoer}, {Larson}, {Papovich}, {Pirzkal},
  {Rojas-Ruiz}, {Song}, {Schneider}, {Somerville}, {Wilkins}, \&
  {Yung}}]{2022ApJ...927..170T}
{Tacchella}, S., {Finkelstein}, S.~L., {Bagley}, M., {et~al.} 2022, \apj, 927,
  170, \dodoi{10.3847/1538-4357/ac4cad}

\bibitem[{{Tremonti} {et~al.}(2004){Tremonti}, {Heckman}, {Kauffmann},
  {Brinchmann}, {Charlot}, {White}, {Seibert}, {Peng}, {Schlegel}, {Uomoto},
  {Fukugita}, \& {Brinkmann}}]{2004ApJ...613..898T}
{Tremonti}, C.~A., {Heckman}, T.~M., {Kauffmann}, G., {et~al.} 2004, \apj, 613,
  898, \dodoi{10.1086/423264}

\bibitem[{{Veilleux} \& {Osterbrock}(1987)}]{1987ApJS...63..295V}
{Veilleux}, S., \& {Osterbrock}, D.~E. 1987, \apjs, 63, 295,
  \dodoi{10.1086/191166}

\bibitem[{{Verhamme} {et~al.}(2017){Verhamme}, {Orlitov{\'a}}, {Schaerer},
  {Izotov}, {Worseck}, {Thuan}, \& {Guseva}}]{2017A&A...597A..13V}
{Verhamme}, A., {Orlitov{\'a}}, I., {Schaerer}, D., {et~al.} 2017, \aap, 597,
  A13, \dodoi{10.1051/0004-6361/201629264}

\bibitem[{{Yang} {et~al.}(2017{\natexlab{a}}){Yang}, {Malhotra}, {Rhoads}, \&
  {Wang}}]{2017ApJ...847...38Y}
{Yang}, H., {Malhotra}, S., {Rhoads}, J.~E., \& {Wang}, J. 2017{\natexlab{a}},
  \apj, 847, 38, \dodoi{10.3847/1538-4357/aa8809}

\bibitem[{{Yang} {et~al.}(2017{\natexlab{b}}){Yang}, {Malhotra}, {Gronke},
  {Rhoads}, {Leitherer}, {Wofford}, {Jiang}, {Dijkstra}, {Tilvi}, \&
  {Wang}}]{2017ApJ...844..171Y}
{Yang}, H., {Malhotra}, S., {Gronke}, M., {et~al.} 2017{\natexlab{b}}, \apj,
  844, 171, \dodoi{10.3847/1538-4357/aa7d4d}

\bibitem[{{Yao} {et~al.}(2022){Yao}, {Liu}, {Kong}, {Gao}, {Chen}, {Chen},
  {Liang}, {Lin}, {Tang}, \& {Zhang}}]{2022ApJ...926...57Y}
{Yao}, Y., {Liu}, H., {Kong}, X., {et~al.} 2022, \apj, 926, 57,
  \dodoi{10.3847/1538-4357/ac3b55}

\bibitem[{{York} {et~al.}(2000){York}, {Adelman}, {Anderson}, {Anderson},
  {Annis}, {Bahcall}, {Bakken}, {Barkhouser}, {Bastian}, {Berman}, {Boroski},
  {Bracker}, {Briegel}, {Briggs}, {Brinkmann}, {Brunner}, {Burles}, {Carey},
  {Carr}, {Castander}, {Chen}, {Colestock}, {Connolly}, {Crocker}, {Csabai},
  {Czarapata}, {Davis}, {Doi}, {Dombeck}, {Eisenstein}, {Ellman}, {Elms},
  {Evans}, {Fan}, {Federwitz}, {Fiscelli}, {Friedman}, {Frieman}, {Fukugita},
  {Gillespie}, {Gunn}, {Gurbani}, {de Haas}, {Haldeman}, {Harris}, {Hayes},
  {Heckman}, {Hennessy}, {Hindsley}, {Holm}, {Holmgren}, {Huang}, {Hull},
  {Husby}, {Ichikawa}, {Ichikawa}, {Ivezi{\'c}}, {Kent}, {Kim}, {Kinney},
  {Klaene}, {Kleinman}, {Kleinman}, {Knapp}, {Korienek}, {Kron}, {Kunszt},
  {Lamb}, {Lee}, {Leger}, {Limmongkol}, {Lindenmeyer}, {Long}, {Loomis},
  {Loveday}, {Lucinio}, {Lupton}, {MacKinnon}, {Mannery}, {Mantsch}, {Margon},
  {McGehee}, {McKay}, {Meiksin}, {Merelli}, {Monet}, {Munn}, {Narayanan},
  {Nash}, {Neilsen}, {Neswold}, {Newberg}, {Nichol}, {Nicinski}, {Nonino},
  {Okada}, {Okamura}, {Ostriker}, {Owen}, {Pauls}, {Peoples}, {Peterson},
  {Petravick}, {Pier}, {Pope}, {Pordes}, {Prosapio}, {Rechenmacher}, {Quinn},
  {Richards}, {Richmond}, {Rivetta}, {Rockosi}, {Ruthmansdorfer}, {Sandford},
  {Schlegel}, {Schneider}, {Sekiguchi}, {Sergey}, {Shimasaku}, {Siegmund},
  {Smee}, {Smith}, {Snedden}, {Stone}, {Stoughton}, {Strauss}, {Stubbs},
  {SubbaRao}, {Szalay}, {Szapudi}, {Szokoly}, {Thakar}, {Tremonti}, {Tucker},
  {Uomoto}, {Vanden Berk}, {Vogeley}, {Waddell}, {Wang}, {Watanabe},
  {Weinberg}, {Yanny}, {Yasuda}, \& {SDSS Collaboration}}]{2000AJ....120.1579Y}
{York}, D.~G., {Adelman}, J., {Anderson}, John~E., J., {et~al.} 2000, \aj, 120,
  1579, \dodoi{10.1086/301513}

\bibitem[{{Yung} {et~al.}(2021){Yung}, {Somerville}, {Finkelstein},
  {Hirschmann}, {Dav{\'e}}, {Popping}, {Gardner}, \&
  {Venkatesan}}]{2021MNRAS.508.2706Y}
{Yung}, L.~Y.~A., {Somerville}, R.~S., {Finkelstein}, S.~L., {et~al.} 2021,
  \mnras, 508, 2706, \dodoi{10.1093/mnras/stab2761}

\bibitem[{{Yung} {et~al.}(2020){Yung}, {Somerville}, {Finkelstein}, {Popping},
  {Dav{\'e}}, {Venkatesan}, {Behroozi}, \& {Ferguson}}]{2020MNRAS.496.4574Y}
---. 2020, \mnras, 496, 4574, \dodoi{10.1093/mnras/staa1800}

\end{thebibliography}
\bibliographystyle{aasjournal}

\appendix

\section{Properties of our samples}

\startlongtable
\begin{deluxetable*}{ccccccccc}
\tabletypesize{\footnotesize}
\tablewidth{0pt} 
\tablecaption{Properties of our compact galaxies with strong [$\mathrm{O \, {\scriptstyle III}}$]$\lambda$5007 emission.
\label{tab:Properties}}
\tablehead{
\colhead{SDSS ID} & \colhead{Ra}& \colhead{Dec} & \colhead{Redshift} &  \colhead{log$L(H\alpha)$} &
\colhead{EW([$\mathrm{O \, {\scriptstyle III}}$])} & \colhead{SFR} & \colhead{log$(M_*/M_{\odot})$} & \colhead{12+log(O/H)} \\
\colhead{} & \colhead{(J2000)}& \colhead{(J2000)} & \colhead{} & \colhead{($\mathrm{erg \, s^{-1}}$)} &
\colhead{($\mathrm{\AA}$)} & \colhead{($\mathrm{M_{\odot} \, yr^{-1}}$)} & \colhead{} & \colhead{}
}
\colnumbers
\startdata 
SDSS J081136.31+304311.1  &  122.90131  &  30.71977  &  0.1575 & 41.83 &  878.98  &  5.31$\pm$0.29  &  8.11$\pm$0.09  &  8.06$\pm$0.25\\
SDSS J164348.57+343935.6  &  250.95239  &  34.65991  &  0.1102 & 40.98  &  921.21  &  0.75$\pm$0.01  &  7.82$\pm$0.13  &  7.78$\pm$0.43\\
SDSS J013425.48+084939.8  &  23.60620  &  8.82772  &  0.0518  &  40.38 & 1132.45  &  0.19$\pm$0.00  &  6.80$\pm$0.17  &  7.81$\pm$0.38\\
SDSS J011720.17+073145.0  &  19.33406  &  7.52919  &  0.2818  & 41.98 & 2223.96  &  7.62$\pm$0.09  &  7.71$\pm$0.09  &  7.91$\pm$0.28\\
SDSS J084022.58+214109.5  &  130.09411  &  21.68599  &  0.1375  & 41.56 &  451.49  &  2.85$\pm$0.03  &  8.51$\pm$0.16  &  7.94$\pm$0.24\\
SDSS J022635.74-011021.5  &  36.64893  &  -1.17264  &  0.0962  &  41.38 & 212.92  &  1.89$\pm$0.04  &  8.72$\pm$0.17  &  8.10$\pm$0.21\\
SDSS J080942.74+491821.1  &  122.42809  &  49.30587  &  0.0781  & 41.03  & 626.04  &  0.85$\pm$0.05  &  8.18$\pm$0.09  &  7.96$\pm$0.25\\
SDSS J083745.20+375237.8  &  129.43834  &  37.87718  &  0.1615  & 41.53  & 622.60  &  2.65$\pm$0.06  &  8.73$\pm$0.14  &  8.23$\pm$0.21\\
SDSS J112624.43-005449.0  &  171.60182  &  -0.91362  &  0.0772  & 40.74 &  690.69  &  0.44$\pm$0.00  &  7.37$\pm$0.02  &  7.77$\pm$0.40\\
SDSS J151806.17+264617.8  &  229.52574  &  26.77163  &  0.0829  & 41.22 &  706.35  &  1.30$\pm$0.02  &  7.93$\pm$0.12  &  8.19$\pm$0.19\\
SDSS J155239.84+170851.2  &  238.16603  &  17.14758  &  0.0721  & 40.62  & 993.21  &  0.32$\pm$0.01  &  6.84$\pm$0.18  &  7.77$\pm$0.38\\
SDSS J153126.82+203027.4  &  232.86176  &  20.50763  &  0.0657 & 40.31 &  408.72  &  0.16$\pm$0.00  &  7.56$\pm$0.10  &  7.93$\pm$0.34\\
SDSS J152826.52+231843.1  &  232.11053  &  23.31198  &  0.0595  & 41.27 &  255.62  &  1.47$\pm$0.02  &  7.88$\pm$0.07  &  7.76$\pm$0.24\\
SDSS J234452.01-005130.0  &  356.21674  &  -0.85835  &  0.3003  & 42.34 & 327.01  &  17.12$\pm$0.14  &  9.33$\pm$0.08  &  8.20$\pm$0.20\\
SDSS J011755.46-001249.5  &  19.48110  &  -0.21375  &  0.1923  & 41.99 & 267.44  &  7.64$\pm$0.07  &  9.43$\pm$0.14  &  8.28$\pm$0.20\\
SDSS J025349.27+030100.6  &  43.45530  &  3.01685  &  0.0724  & 40.62 & 996.21  &  0.33$\pm$0.00  &  7.45$\pm$0.16  &  7.92$\pm$0.40\\
SDSS J234916.37+043000.1  &  357.31822  &  4.50003  &  0.0667  & 40.37 & 880.01  &  0.18$\pm$0.00  &  7.01$\pm$0.12  &  7.87$\pm$0.42\\
SDSS J224750.52+040721.4  &  341.96051  &  4.12262  &  0.0378  & 40.30 & 375.70  &  0.16$\pm$0.00  &  6.86$\pm$0.14  &  7.91$\pm$0.29\\
SDSS J083516.74+114122.0  &  128.81979  &  11.68945  &  0.0583  & 40.49 & 780.82  &  0.24$\pm$0.00  &  7.04$\pm$0.11  &  7.91$\pm$0.32\\
SDSS J075321.62+095344.0  &  118.34012  &  9.89556  &  0.0719  & 40.67 & 1617.05  &  0.37$\pm$0.00  &  7.07$\pm$0.16  &  7.75$\pm$0.37\\
SDSS J112945.62+355107.0  &  172.44012  &  35.85195  &  0.0561  & 40.40 & 426.83  &  0.20$\pm$0.00  &  7.04$\pm$0.11  &  7.68$\pm$0.45\\
SDSS J120557.30+025656.5  &  181.48878  &  2.94904  &  0.0766  & 40.81 & 346.29  &  0.51$\pm$0.01  &  8.01$\pm$0.17  &  8.03$\pm$0.25\\
SDSS J114653.25+034241.9  &  176.72189  &  3.71165  &  0.0447  & 40.66 & 351.80  &  0.36$\pm$0.00  &  7.41$\pm$0.11  &  8.07$\pm$0.24\\
SDSS J111547.88+071910.2  &  168.94953  &  7.31952  &  0.0619  & 40.50 & 571.16  &  0.25$\pm$0.01  &  7.11$\pm$0.11  &  8.01$\pm$0.27\\
SDSS J224953.77+164659.9  &  342.47407  &  16.78331  &  0.0882  & 40.56 & 868.13  &  0.28$\pm$0.01  &  7.44$\pm$0.12  &  8.04$\pm$0.29\\
SDSS J225113.60+095512.4  &  342.80668  &  9.92013  &  0.0332  & 39.86 & 359.21  &  0.06$\pm$0.00  &  7.01$\pm$0.19  &  7.81$\pm$0.36\\
SDSS J161221.60+110516.4  &  243.09003  &  11.08790  &  0.0722  & 41.32 & 654.65  &  1.66$\pm$0.02  &  8.01$\pm$0.16  &  8.11$\pm$0.20\\
SDSS J120402.17+072558.6  &  181.00906  &  7.43296  &  0.0322  & 40.07 & 214.85  &  0.09$\pm$0.00  &  6.38$\pm$0.10  &  8.08$\pm$0.22\\
SDSS J124018.08+130344.5  &  190.07535  &  13.06237  &  0.0632  & 40.46 & 249.97  &  0.23$\pm$0.00  &  8.20$\pm$0.13  &  7.98$\pm$0.28\\
SDSS J140153.86+123349.8  &  210.47442  &  12.56386  &  0.0619  & 40.84 & 475.77  &  0.54$\pm$0.02  &  7.52$\pm$0.16  &  8.00$\pm$0.28\\
SDSS J145425.33+121221.6  &  223.60555  &  12.20601  &  0.1508  & 41.67 & 278.82  &  3.72$\pm$0.09  &  7.89$\pm$0.17  &  7.82$\pm$0.37\\
SDSS J000441.90+132012.4  &  1.17459  &  13.33678  &  0.0548  & 40.61 & 876.37  &  0.32$\pm$0.00  &  7.02$\pm$0.12  &  7.74$\pm$0.31\\
SDSS J091819.28+305654.8  &  139.58036  &  30.94858  &  0.0946  & 41.43 & 728.39  &  2.15$\pm$0.04  &  8.21$\pm$0.12  &  7.99$\pm$0.22\\
SDSS J143251.79+271111.2  &  218.21581  &  27.18645  &  0.0609  & 40.40 & 305.83  &  0.20$\pm$0.00  &  7.54$\pm$0.16  &  8.03$\pm$0.25\\
SDSS J231521.95+214328.3  &  348.84148  &  21.72454  &  0.0549  & 40.88 & 1218.94  &  0.60$\pm$0.01  &  6.93$\pm$0.13  &  7.91$\pm$0.21\\
SDSS J004954.52+285449.0  &  12.47719  &  28.91362  &  0.0390  & 40.10  & 535.40  &  0.10$\pm$0.00  &  6.87$\pm$0.14  &  7.77$\pm$0.37\\
SDSS J004641.11+300259.1  &  11.67133  &  30.04976  &  0.0426  & 40.16 & 513.58  &  0.11$\pm$0.00  &  7.01$\pm$0.12  &  7.94$\pm$0.33\\
SDSS J101028.98+231259.0  &  152.62078  &  23.21641  &  0.0664  & 41.41 & 556.81  &  2.01$\pm$0.04  &  8.01$\pm$0.13  &  7.97$\pm$0.21\\
SDSS J115907.58+292810.8  &  179.78162  &  29.46967  &  0.0511  & 40.51 & 503.12  &  0.25$\pm$0.01  &  7.13$\pm$0.09  &  7.97$\pm$0.26\\
SDSS J142207.05+505757.1  &  215.52941  &  50.96588  &  0.1813  & 42.21 & 882.05  &  12.76$\pm$0.14  &  8.58$\pm$0.13  &  8.29$\pm$0.19\\
SDSS J154507.90+574702.5  &  236.28294  &  57.78404  &  0.0751  & 40.54 & 688.18  &  0.27$\pm$0.00  &  7.14$\pm$0.09  &  7.98$\pm$0.31\\
SDSS J125930.29+561910.7  &  194.87624  &  56.31966  &  0.0721  & 40.56 & 1995.34  &  0.29$\pm$0.01  &  6.48$\pm$0.07  &  7.49$\pm$0.71\\
SDSS J013411.59-051747.9  &  23.54832  &  -5.29665  &  0.0951  & 41.67 & 648.34  &  3.73$\pm$0.04  &  8.60$\pm$0.16  &  8.22$\pm$0.20\\
SDSS J093355.08+510924.6  &  143.47951  &  51.15683  &  0.2913  & 42.19 & 1661.73  &  12.29$\pm$0.13  &  7.98$\pm$0.08  &  7.96$\pm$0.27\\
SDSS J091216.82+482004.8  &  138.07011  &  48.33469  &  0.3063  & 42.08 & 735.10  &  9.55$\pm$0.13  &  8.75$\pm$0.17  &  8.13$\pm$0.24\\
SDSS J112652.66+460501.3  &  171.71944  &  46.08371  &  0.3476  & 42.16 & 395.36  &  11.33$\pm$0.13  &  8.94$\pm$0.13  &  8.00$\pm$0.28\\
SDSS J124113.49+491143.2  &  190.30621  &  49.19536  &  0.1439  & 41.45 & 693.48  &  2.21$\pm$0.03  &  8.07$\pm$0.13  &  8.07$\pm$0.29\\
SDSS J085503.74+471818.5  &  133.76561  &  47.30515  &  0.1585  & 42.31 & 564.37  &  16.22$\pm$0.85  &  8.44$\pm$0.13  &  8.00$\pm$0.32\\
SDSS J221042.80+184324.2  &  332.67834  &  18.72339  &  0.0784  & 40.63 & 389.13  &  0.34$\pm$0.01  &  8.04$\pm$0.12  &  8.00$\pm$0.26\\
SDSS J004213.42+231047.2  &  10.55592  &  23.17978  &  0.0575  & 40.50 & 217.95  &  0.25$\pm$0.00  &  7.18$\pm$0.24  &  7.93$\pm$0.27\\
SDSS J004221.14+252353.1  &  10.58809  &  25.39808  &  0.1438  & 41.25 & 256.71  &  1.40$\pm$0.04  &  8.54$\pm$0.18  &  7.85$\pm$0.37\\
SDSS J014137.14+275018.5  &  25.40477  &  27.83850  &  0.0586  & 40.58 & 1033.71  &  0.30$\pm$0.00  &  6.45$\pm$0.12  &  7.91$\pm$0.38\\
SDSS J000115.81+270025.1  &  0.31589  &  27.00699  &  0.0922  & 40.31 & 1962.18  &  0.16$\pm$0.02  &  7.15$\pm$0.08  &  8.28$\pm$0.27\\
SDSS J233648.13+264338.5  &  354.20056  &  26.72737  &  0.1642  & 41.50 & 946.34  &  2.47$\pm$0.04  &  8.00$\pm$0.11  &  7.99$\pm$0.24\\
SDSS J231123.10+245547.3  &  347.84628  &  24.92981  &  0.1454  & 41.60 & 369.21  &  3.12$\pm$0.02  &  8.34$\pm$0.20  &  8.09$\pm$0.22\\
SDSS J221955.10+303036.6  &  334.97959  &  30.51017  &  0.1760  & 41.56 & 825.83  &  2.88$\pm$0.04  &  7.92$\pm$0.08  &  7.90$\pm$0.27\\
SDSS J021442.13+313608.0  &  33.67555  &  31.60223  &  0.0538  & 39.92 &  703.59  &  0.07$\pm$0.00  &  6.69$\pm$0.21  &  7.88$\pm$0.43\\
SDSS J023526.26+020638.2  &  38.85943  &  2.11061  &  0.0220  & 40.46 & 251.63  &  0.23$\pm$0.01  &  5.85$\pm$0.15  &  8.17$\pm$0.21\\
SDSS J013808.83-001933.6  &  24.53682  &  -0.32600  &  0.0562  & 41.15 &  420.95  &  1.11$\pm$0.01  &  6.68$\pm$0.08  &  8.15$\pm$0.21\\
SDSS J112608.24+575531.8  &  171.53436  &  57.92553  &  0.1807  & 42.05  & 370.44  &  8.87$\pm$0.11  &  8.84$\pm$0.13  &  8.19$\pm$0.20\\
SDSS J135716.14+555310.4  &  209.31727  &  55.88624  &  0.0373  & 40.09 & 644.94  &  0.10$\pm$0.00  &  6.51$\pm$0.15  &  8.06$\pm$0.28\\
SDSS J135613.67+521945.5  &  209.05699  &  52.32931  &  0.1115  & 41.08  & 1877.47  &  0.95$\pm$0.03  &  7.27$\pm$0.17  &  7.92$\pm$0.41\\
SDSS J132116.78+513926.8  &  200.31992  &  51.65745  &  0.1517  & 41.86 & 277.89  &  5.72$\pm$0.08  &  8.77$\pm$0.11  &  8.20$\pm$0.20\\
SDSS J125701.58+595401.0  &  194.25662  &  59.90030  &  0.0590  & 40.40 & 388.64  &  0.20$\pm$0.00  &  7.38$\pm$0.12  &  7.99$\pm$0.28\\
SDSS J130735.50+520221.7  &  196.89796  &  52.03939  &  0.1523  & 42.03 & 949.86  &  8.55$\pm$0.13  &  7.85$\pm$0.10  &  7.70$\pm$0.35\\
SDSS J075742.06+451846.2  &  119.42526  &  45.31285  &  0.0546  & 41.04 & 438.05  &  0.86$\pm$0.01  &  7.06$\pm$0.07  &  8.28$\pm$0.19\\
SDSS J123548.35+434221.5  &  188.95150  &  43.70599  &  0.3197  & 42.37 & 847.41  &  18.38$\pm$0.53  &  8.51$\pm$0.13  &  8.27$\pm$0.20\\
SDSS J131829.29+444909.1  &  199.62205  &  44.81920  &  0.1574  & 41.59 & 284.00  &  3.05$\pm$0.04  &  8.64$\pm$0.11  &  8.13$\pm$0.22\\
SDSS J132653.76+435741.9  &  201.72401  &  43.96167  &  0.1386  & 41.52 & 363.34  &  2.60$\pm$0.07  &  8.43$\pm$0.14  &  8.07$\pm$0.22\\
SDSS J154509.39+503448.8  &  236.28915  &  50.58024  &  0.0573  &  40.70 & 211.47  &  0.40$\pm$0.01  &  8.02$\pm$0.16  &  8.05$\pm$0.25\\
SDSS J142053.29+575442.7  &  215.22206  &  57.91187  &  0.0534  & 40.43 & 359.26  &  0.21$\pm$0.00  &  7.31$\pm$0.16  &  7.97$\pm$0.25\\
SDSS J144359.00+462106.4  &  220.99587  &  46.35179  &  0.1496  & 41.17 & 412.84  &  1.16$\pm$0.01  &  8.01$\pm$0.17  &  7.87$\pm$0.34\\
SDSS J152352.63+425157.0  &  230.96931  &  42.86583  &  0.3260  & 42.19 & 656.16  &  12.38$\pm$0.12  &  8.74$\pm$0.17  &  8.13$\pm$0.24\\
SDSS J163845.33+412432.9  &  249.68889  &  41.40916  &  0.0727  & 40.68 & 345.42  &  0.38$\pm$0.00  &  6.67$\pm$0.12  &  8.10$\pm$0.23\\
SDSS J121823.93+392508.7  &  184.59971  &  39.41910  &  0.0510  & 40.68 & 1120.17  &  0.38$\pm$0.01  &  6.60$\pm$0.09  &  7.91$\pm$0.25\\
SDSS J124845.07+401913.8  &  192.18782  &  40.32053  &  0.0716  & 40.66 & 1157.42  &  0.36$\pm$0.00  &  6.49$\pm$0.05  &  7.88$\pm$0.40\\
SDSS J133107.26+363337.9  &  202.78029  &  36.56054  &  0.1143  & 41.02 & 1402.93  &  0.82$\pm$0.01  &  7.25$\pm$0.15  &  7.80$\pm$0.40\\
SDSS J212827.39+000822.3  &  322.11413  &  0.13953  &  0.1645  & 41.98 & 1172.76  &  7.55$\pm$0.15  &  7.85$\pm$0.03  &  7.89$\pm$0.26\\
\enddata
\tablecomments{(1) object name; (2-3) right ascension and declination in J2000 from SDSS DR16; (4) SDSS spectroscopic redshift; (5) logarithmic H$\alpha$ luminosity in erg s$^{-1}$. (6) [$\mathrm{O \, {\scriptstyle III}}$] equivalent width in \AA; (7) Star formation rate estimated from H$\alpha$ in $M_{\odot} \mathrm{yr}^{-1}$; (8) logarithmic stellar mass in $M_{\odot}$. (9) metallicity estimated with the [$\mathrm{N \, {\scriptstyle II}}$] line}
\end{deluxetable*}

\end{document}